%% file: 00-main.tex
\documentclass[acmsmall]{acmart}

\input{01-packages}

\begin{document}

\input{02-top-matter}
\input{03-title-authors}

\input{04-abstract}
\input{05-bottom-matter}

\input{10-intro}

\input{20-related-works}
\input{30-proposed-approach}

\input{40-examples}

\input{41-virtue.tex}

\input{42-graphprints.tex}

\input{90-conclusion}

\input{99-acks} 

\small

Bibliography

\input{refs.bbl}
\end{document}

%% file: 01-packages.tex
\usepackage{booktabs} 

\usepackage{hyperref}

\usepackage{lipsum}
\usepackage{courier} 
\usepackage{tabularx}
\usepackage{graphicx}
\graphicspath{ {./images/} } 
\usepackage{svg}
\usepackage{pdflscape}
\usepackage{makecell}
\usepackage{natbib} 
\usepackage{enumitem} 
\usepackage[flushleft]{threeparttable}
\usepackage{multirow}

\usepackage{todonotes} 
\usepackage{wrapfig}

\usepackage[ruled]{algorithm2e} 
\usepackage{xcolor,colortbl}
\usepackage{ragged2e}

\definecolor{LightCyan}{RGB}{112,172,203}
\definecolor{LightGrey}{RGB}{238,238,238}

\SetAlFnt{\small}
\SetAlCapFnt{\small}
\SetAlCapNameFnt{\small}
\SetAlCapHSkip{0pt}
\IncMargin{-\parindent}

\usepackage{soul}
\usepackage{color}

\usepackage{capt-of}
\usepackage{booktabs} 
\usepackage{arydshln} 
\usepackage[flushleft]{threeparttable}

%% file: 02-top-matter.tex
\acmJournal{CSUR}
\acmVolume{9}
\acmNumber{4}
\acmArticle{39}
\acmYear{2018}
\acmMonth{3}
\acmArticleSeq{11}


\setcopyright{usgovmixed}

\acmDOI{0000001.0000001}

\received{February 2007}
\received{March 2009}
\received[accepted]{June 2009}

%% file: 03-title-authors.tex
\title{Quantifiable \& Comparable Evaluations of Cyber Defensive Capabilities: \\
A Survey \& Novel, Unified Approach}

\author{Michael D. Iannacone}
\orcid{0003-3081-4761}
\email{iannaconemd@ornl.gov}
\affiliation{%
  \institution{Cyber \& Applied Data Analytics Division, Oak Ridge National Laboratory}
  \streetaddress{1 Bethel Valley Road}
  \city{Oak Ridge}
  \state{TN}
  \postcode{37831}
  \country{USA}
}

\author{Robert A. Bridges}
\orcid{0001-7962-6329}
\email{bridgesra@ornl.gov}
\affiliation{%
  \institution{Cyber \& Applied Data Analytics Division, Oak Ridge National Laboratory}
  \streetaddress{1 Bethel Valley Road}
  \city{Oak Ridge}
  \state{TN}
  \postcode{37831}
  \country{USA}
}



%% file: 04-abstract.tex
\begin{abstract}
Metrics and frameworks to quantifiably assess security measures have arisen from needs of three distinct research communities\textemdash statistical measures from the intrusion detection and prevention literature, evaluation of  cyber exercises, e.g., red-team and capture-the-flag competitions, and economic analyses addressing cost-versus-security tradeoffs. 
In this paper we provide two primary contributions to the security evaluation literature---a representative survey, and a novel framework for evaluating security that is flexible, applicable to all three use cases, and readily interpretable. In our survey of the literature we identify the distinct themes from each community's evaluation procedures side by side and flesh out the drawbacks and benefits of each. 

The evaluation framework we propose includes comprehensively modeling the resource, labor, and attack costs in dollars incurred based on expected resource usage, accuracy metrics, and time. 
This framework provides a unified approach in that it incorporates 
the accuracy and performance metrics, which dominate intrusion detection evaluation, 
the time to detection and impact to data and resources of an attack, favored by educational competitions' metrics, and the monetary cost of many essential security components used in financial analysis.   
Moreover, it is flexible enough to accommodate each use case, easily interpretable and comparable, and comprehensive in terms of costs considered. 
Finally, we provide two examples of the framework applied to real-world use cases. 
Overall, we provide a survey and a grounded, flexible framework with multiple concrete examples for evaluating security which can address the needs of three currently distinct communities. 

\end{abstract}

%% file: 05-bottom-matter.tex
%
%


%
%


\thanks{This manuscript has been authored by UT-Battelle, LLC, under contract DE-AC05-00OR22725 with the US Department of Energy (DOE). The US government retains and the publisher, by accepting the article for publication, acknowledges that the US government retains a nonexclusive, paid-up, irrevocable, worldwide license to publish or reproduce the published form of this manuscript, or allow others to do so, for US government purposes. DOE will provide public access to these results of federally sponsored research in accordance with the DOE Public Access Plan (http://energy.gov/downloads/doe-public-access-plan).}
\maketitle

\renewcommand{\shortauthors}{Iannacone, Bridges, \& Long}

%% file: 10-intro.tex
\section{Introduction} 

As security breaches continue to affect personal resources, industrial systems, and enterprise networks, there is an ever growing need to understand, ``How secure are my systems?'' 
This need has driven diverse efforts to systematize an answer that provides meaningful quantifiable comparisons. 
In the research literature, evaluation of information security measures has developed from three different but related communities. 
A vibrant and growing body of research on intrusion detection and prevention systems (IDS/IPS) has produced algorithms, software, and system frameworks to increase security; consequently, evaluation criteria to assess the efficacy of these ideas has been informally adopted. 
Most common methods require representative datasets with labeled attacks and seek traditional statistical metrics, such as true/false positive rates.

Analogous developments emerged from cyber security  exercises, which have become commonplace activities for education and for enterprise self assessment. 
Scoring for red-team and capture-the-flag educational exercises are necessary to the framework of the competitions, and generally provide more concrete measures of security as they can accurately quantify measures of network resources, e.g., the time that a server was offline or the number of records stolen. 
In a similar vein to these competitions are red team events and penetration testing of software or systems. 
This practice is commonplace and seeks to enumerate all vulnerabilities in a product or security posture.
While there is a body or research literature on red teaming and penetration testing (independent of cyber competitions), it does not provide quantifiable assessments for security (as is necessary for cyber competitions). 
Rather, it provides specific lists of weaknesses. 
Thus penetration testing is a powerful capability proving important information for hardening and qualitative understanding of security. 

More pragmatic needs for quantifying security arise at the interface of an organization's financial and security management. 
Justifying security budgets and identifying optimal use of resources requires concise but revealing metrics intelligible to security experts and non-experts alike. 
To this end, intersections of the security and economic research communities have developed cost-benefit analyses that give methods to determine the value of security. 
As we shall see, the desired summary statistics provide a platform for quantifiable analysis, but are often dependent on intangible estimates, future projections, and lack semantic understanding.

While these sub-domains of security research developed rather independently, their overarching theme is the same---to evaluate security of systems, operators, and processes. 
In particular, the metrics developed by each lane of research seek to
\textbf{provide a quantifiable, comparable measure to validate and reason about the efficacy of security capabilities.} 
The goal of this article is to build on the advantages of these previous works to provide such a scoring framework that accommodates analysis of security tools and procedures, especially to support security and network operation centers. 
While our focus is admittedly on comparing and evaluating the overall effects of intrusion detection capabilities, the framework can indeed be used for other types of tools or changes in procedures. 
For example, adopting a modern SIEM (security and incident event management system) tool or changing operating procedures may both lower investigation and response times (which is taken into account in the model), and hence can be compared to each other as well as to, say, adoption of an endpoint detection capability that increases accuracy of alerts but not efficiency of response.

This work delivers two primary contributions. 
First, we provide a representative survey of the security evaluation literature (Section \ref{sec:related-work}); that is, those portions of the literature developing quantifiable measures of security capabilities. 
Works are chosen that collectively highlight the trends and landmarks from each subdomain (IDS evaluation, cyber competition scoring, economic cost-benefit analyses)  allowing side-by-side comparison. 
We illustrate drawbacks and beneficial properties of these evaluations techniques. 

Second, we propose a  ``unified'' framework for evaluating security measures by modeling costs of non-attack and attack states.   
By ``unified", we mean this approach  provides the configurability to be comprehensive in terms of the real-world factors that contribute to the  models, and it is flexible enough to satisfy all three use cases. 
Specifically, it incorporates the accuracy and performance, which dominate IDS evaluation; 
the time to detection as well as the confidentiality, integrity, and availability 
of resources, favored by  competitions' metrics;  
and  the dollar costs or resources, labor, and attacks comprised by cost-benefit analyses. 

Our model, described in Section \ref{sec:new_approach}, is a cost model that can be configured for many diverse scenarios, and permits a variety of granularity in modeling each component to accommodate situations with ample/sparse information. 
Unlike many previous frameworks, ours uses a single, easy-to-interpret metric, cost in dollars, and is readily analyzable as each component of this cost uses a fairly simple model. 
As is commonplace for such economic models, finding accurate input values (e.g., maximum possible cost of an attack, or the quantity of false alerts expected) is difficult and a primary drawback of our and all previous similar models (see Section \ref{sec:cost_benefit}). 
In response, we provide a sensitivity analysis is Section \ref{sec:full-model}, to identify the model parameters/components that have the greatest effect, so users know where to target efforts to increase accuracy---a practice that our survey reveals is unfortunately rare. 

We employ the new model in Section \ref{sec:examples}. 
As the driving force behind this research, we give a detailed configuration of our cost model to be used as the evaluation procedure for an upcoming IAPRA Challenge involving intrusion detection (Section \ref{sec:virtue}). 
We provided simulated attack and defense scenarios to test our scoring framework and exhibit results confirming that the evaluation procedure encourages a balance of accuracy, timeliness, and resource costs. 
We expect our simulation work can provide a baseline for future competitors. 

As another example we configure this new model to evaluate the GraphPrints IDS from our previous work \cite{harshaw2016graphprints}. 
This example shows the efficacy of the evaluation model from many viewpoints. 
For researchers it provides an alternative to simple accuracy metrics, by incorporating the accuracy findings and resource costs into a realistic, quantifiable cost framework. 
This allows, for example, optimizing thresholds, accuracy, and performance considerations rather than just reporting each. 
From the point of view of a security operation center (SOC), we provide an example of how to evaluate a tool with the perspective of a potential purchase. 
Finally, we consider the model from a vendor's eye, and derive bounds for the potential licensing costs.


Overall, the main contributions of this work are (1) a survey of three distinct but related areas, and (2) a general framework allowing computation and comparison of security that satisfies the needs of all three use cases with examples of how this metric can be used.

%% file: 20-related-works.tex
\section{Related Work: A Survey of Quantifiable Security Evaluations} 
\label{sec:related-work} 

For related works, we focus on those that develop or use quantifiable means to compare and evaluate security capabilities. 
In this review, we found that evaluation of cyber security measures has developed in three, rather independent threads. 
This section gives a survey of our findings and strives to be representative of the main ideas in each topic area. 
Since we cannot comprehensively cite all related papers, we focus on papers with high impact or which illustrate particularly important developments within their topic area. 

\input{22-ids.tex}
\input{24-competitions.tex}

\input{26-cost-benefit.tex}

%% file: 22-ids.tex
\subsection{Evaluation of Intrusion Detection \& Prevention Systems} 
\label{sec:ids}

In the intrusion detection and prevention research, which focuses on evaluating and comparing detection capabilities, 
researchers generally seek statistical evidence for detection accuracy and computational viability as calculated on test datasets. 
While such evaluations are 
commonplace, curation of convincing test sets and developing relevant metrics for efficacy in real-world use has proven difficult. 
The default metrics employed, such as computational complexity for performance and the usual accuracy metrics (e.g., true/false positive rate, precision, receiver operating characteristic (ROC) curve, and area under the curve (AUC) to name a few) are indeed important statistics to consider. 
Yet, these metrics cannot account for many important  operational considerations, e.g., valuing earlier detection over later, valuing high-priority resources/data, or including the costs of operators' time.   
Research to incorporate these aspects is emerging, e.g.~\cite{garcia2014empirical}, but still does not incorporate all of these real-world concerns. 
Often the cost to implement the proposed security measures in operations, e.g., hardships of training the algorithms on network-specific data or configuration/reconfiguration costs, is either neglected or considered out of scope.  

Further,  validation of a proposed method requires data with known attacks and enough fidelity to demonstrate the method's abilities. 
In general, there is limited 
availability of real network datasets with known attacks, and there is little agreement on what qualities  constitute a ``good'', that is, representative and realistic, dataset.  
This is exacerbated by privacy concerns that inhibit releasing real data, and unique characteristics inherent to each network, which limits generalizability of any given dataset. 
Notably, there are a small number of publicly available datasets, that have catalyzed a large body of detection research, in spite of many of these datasets receiving ample criticism. 
Notably, the Canadian Institute for Cybersecurity provides 13 datasets that are useful for meaningful, head-to-head validation \url{https://www.unb.ca/cic/datasets/index.html}. 
All of these data sets and many others from the research literature appear in our related survey of host-based IDS works \cite{glass2018survey}. 
In other cases, researchers often use one-off, custom-made datasets, e.g.~\cite{jewell2011host, harshaw2016graphprints}. 
While this can potentially address some of the concerns above, these datasets generally are not made publicly available, 
which sacrifices reproducibility of results, inhibits meaningful comparison across publications, 
and  makes it difficult or impossible to verify the quality of these datasets.


\subsubsection{DARPA Dataset} 

One of the earliest attempts to systematize this performance evaluation was the DARPA 1998 dataset, which was originally used to evaluate performers in the 1998 DARPA/AFRL 
``Intrusion Detection Evaluation,'' a \\
competition-style project.
This dataset included both network and host data, including \texttt{tcpdump} and \texttt{list} files, as well as BSM Audit data.  The test network included thousands of simulated machines, generating realistic traffic in a variety of services, over a duration of several weeks; this also included hundreds of attacks of 32 attack types~\cite{lippmann2000evaluating}.
Interestingly, results were measured in ``false alarms per day'' (presumably the averaged per day in the dataset) 
not the raw false alarm rate (number of false positives / number of negatives).  
This false alarm per day rate was compared with the true positive rate in percent. 
The authors chose this representation of the false alarm rate to emphasize the costs in terms of analysts' time; 
however, future researchers generally used the raw false alarm rate instead, since the two are directly proportional.

This was apparently the first use of the ROC curve\footnote{Receiver Operating Characteristic (ROC) refers to the curve plotting the true positive versus false positive rate as the detection threshold is varied.}  in intrusion detection, 
which has since become a common practice.  
As the authors explain:
\begin{quote}
ROC curves for intrusion detection indicate how the detection rate changes as internal
thresholds are varied to generate more or fewer false alarms to
tradeoff detection accuracy against analyst workload.
Measuring the detection rate alone only indicates the types of
attacks that an intrusion detection system may detect.
Such measurements do not indicate the human workload required to
analyze false alarms generated by normal background traffic.
False alarm rates above hundreds per day make a system almost unusable,
even with high detection accuracy, because putative
detections or alerts generated can not be believed and security
analysts must spend many hours each day dismissing false alarms.
Low false alarm rates combined with high detection rates, however,
mean that the putative detection outputs can be
trusted and that the human labor required to confirm detections is minimized.
\end{quote}

This dataset has faced various criticism for not being representative of real-world network conditions.
Some researchers have observed artifacts of the data creation, due to the simulated environment used to generate this data, and explained how these artifacts could bias any detection metrics~\cite{mahoney2003analysis}. 
Other researchers noted that high-visibility but low-impact probing and DoS attacks made up a large proportion of the attacks in the dataset~\cite{brugger2007assessment}, giving them increased importance in the scoring, while other work has criticized the attack taxonomy and scoring methodology~\cite{mchugh2000testing}. 


\subsubsection{KDD Cup 1999 dataset \& evaluation}

The KDD Cup competition was based on the same original dataset as DARPA 98,
but modified to include only the network-oriented features, and pre-processed these into convenient collections of feature vectors. 
This resulted in a dataset consisting essentially of flow data with some additional annotations.
This simplified dataset was easier for researchers to use 
and spawned a large variety of works which applied existing machine learning techniques to this dataset~\cite{glass2018survey}.

In the performer evaluation, the scores were determined by finding the confusion matrix (number of true/false positives/negatives) and multiplying each cell by a factor between 1 and 4. 
This weighting did penalize false positives for the ``user to root'' and ``remote to local'' attack categories more heavily than other types of mis-categorization;
however, this was apparently to compensate for the uneven sizes of the classes,
and did not seem to consider any relative, real-world costs of false positives versus false negatives, 
as these costs were discussed in neither the task description nor the evaluation discussion.
Other than these weights, the evaluation and the discussion of results placed
no particular emphasis on the impact to the operator, 
like the DARPA 98 evaluation discussed above.
Omitting any ROC curves also seemed to be a step backwards, although understandable because not all submissions included a tunable parameter that this requires.

Overall, the performers achieved reasonable results, but none were overwhelmingly successful.
Notably, some very simple methods (e.g., nearest neighbor classifier) performed nearly as well as the winning entries \cite{elkan2000results}.

Because it was derived from the DARPA dataset above, this dataset inherited many of the same problems discussed earlier.
In particular, Sabhnani \& Serpen~\cite{sabhnani2004machine} note that it was very difficult for performers to classify the user-to-root and
remote-to-local categories:
\begin{quote}
Analysis results clearly suggest
that no pattern classification or machine learning algorithm can be trained
successfully with the KDD data set to perform misuse detection for user-to-root or
remote-to-local attack categories.
\end{quote}

These authors explain that the training and testing sets are too different (for these categories)
for any machine learning approach to be effective. 
After merging the training and test sets, they re-tested using five-fold cross validation to observe that the same methods on this modified dataset resulted in vastly superior detection performance. 


Other researchers released variants of this dataset to address some of these problems, such as the NSL-KDD dataset~\cite{tavallaee2009detailed}, which removed many redundant flows and created more balanced classes.  
However, other problems still remain, and at this point the normal traffic and the attacks are no longer representative of modern networks.

\subsubsection{Later Developments}
Following the release of these early benchmark datasets, researchers have generally focused on creating more recent and/or higher-quality datasets, 
and generally have not focused on the methodology/metrics used in evaluation.  Examples are the UNM dataset~\cite{UNM} of system call traces for specific processes, the ADFA datasets~\cite{creech2013generation} of Windows and Linux host audit data, and the VAST competition 2012\footnote{http://www.vacommunity.org/VAST+Challenge+2012} and 2013\footnote{http://www.vacommunity.org/VAST+Challenge+2013} datasets, which focus on various network data sources.

There are many additional data sources which are now publicly available, but most are not suitable as-is for training and testing these systems.  Some are more specialized datasets, such as the Active DNS Project~\cite{kountouras2016enabling}, many contain only malicious traffic, and some data sources are both specialized and malicious, such as only containing peer-to-peer botnet command and control traffic.

There are much fewer examples of datasets of normal traffic, although a few of them have been released, such as CAIDA's anonymized internet traces\footnote{The CAIDA UCSD Anonymized Internet Traces 
\url{http://www.caida.org/data/passive/passive_dataset.xml}}.
These are generally either anonymized and/or aggregated in some way (eg. flows versus packets, modifying addresses, etc.) that can limit some types of analysis.
More importantly, there is little consensus on what ``normal traffic'' means, or what datasets would be representative of what networks.  
If these systems are deployed on networks with different characteristics than the evaluation datasets, the performance could differ significantly. 
Some works have proposed criteria which high-quality datasets should strive to achieve, but these are not universally accepted currently~\cite{sharafaldin2017towards}.

Because of these issues, current datasets can vary significantly in quality, and most are flawed in some way.  
There is currently no consensus on any  general-purpose benchmark datasets, the role which the DARPA and KDD datasets used to fill.  
This situation does seem to be gradually improving over time, but for now these remain significant issues.

\subsubsection{CTU Botnet Dataset \& Time-Dependent Accuracy Metrics}
\label{sec:ctu} 
One notable recent contribution presents both new \\ datasets as well as a new methodology for evaluating IDS performance for detecting botnet traffic from PCAPs\footnote{Captures of network packets} or network flows\footnote{Network flow data, or flows, are the meta-data of network communications, including but not limited to the source and destination IPs, source and destination ports, protocol, timestamp, and quantities of information (bytes, packets, etc.) sent in each direction.}~\cite{garcia2014empirical}.  
The datasets contain a variety of individual malware PCAP files, collected over long time frames, and also some collections of (real) background traffic from a university network. 
In addition, the authors discuss some important shortcomings of the generally-used metrics, and propose some significant improvements.
\begin{quote}
The classic error metrics were defined from a statistical point of view, and they fail to address the detection needs of a network administrator.
\end{quote}
To address this, they propose new criteria. 
For the accuracy metrics, the approach is to break the data into time-windows  and define a confusion matrix (counts of true/false positives/negatives) for each time-window, with increased emphasis for early detection. 
They define 
\begin{equation}
\mbox{correcting\_function}_\alpha (i) = e^{(-\alpha i)}+1
\end{equation}
where $i$ is the number of the time window, and $\alpha$ is an adjustable time-scaling parameter.  
This correcting function is used to define the time-dependent true positive $TP(i)$, false negative $FN(i)$, true negative $TN(i)$, and false positive $FP(i)$ quantities, defined below.  
In what follows, $N_b(i)$ is the number of unique botnet IP addresses 
and $N_n(i)$ is the number of unique normal IP addresses in the  time frame $i$.

\begin{itemize} 
    \item Performance should be measured by addresses instead of flows. This is important because, for example, one malware sample may generate much more command and control traffic than another, even while performing similar actions. 
    This incidental difference in behavior should not artificially impact the detection scores.
    
    \item When correctly detecting botnet traffic, a True Positive, early detection is better than latter.
    \begin{itemize} 
        \item The number of true positives in time window $i$ is $C_{TP}(i)$. 
        They define ``A True Positive is accounted when a Botnet IP address is detected as Botnet at least once during the comparison time frame.''
        \item The time-dependent quantity representing true positives  in time window $i$ is
        $$TP(i) = \frac{C_{TP} (i) (1+e^{(-\alpha i)})}{N_b(i)}$$
    \end{itemize}

    \item When failing to detect actual botnet traffic, a False Negative, an early miss is worse than latter.
    \begin{itemize} 
        \item The number of false negatives occurring in time window $i$ is   $C_{FN}(i).$ 
        They define,  ``A False Negative is accounted when a Botnet IP address is detected as Non-Botnet during the whole comparison time frame.''
        \item The time-dependent quantity representing false negatives  in time window $i$ is 
        $$FN(i) = \frac{C_{FN}(i) (1+e^{(-\alpha i)})}{N_b(i)}.$$
    \end{itemize}

    \item The value of correctly labeling the non-botnet traffic (True Negative) is not affected by time.
    \begin{itemize} 
        \item The number of true negatives occurring in time window $i$ is $C_{TN}(i)$. 
    They define ``A True Negative is accounted when a Normal IP address is detected as Non-Botnet during the whole comparison time frame.''
        \item The time-dependent quantity representing true negatives  in time window $i$ is 
        $$TN(i) = \frac{C_{TN}(i)}{N_n(i)}.$$
    \end{itemize}

    \item The value of incorrectly alerting on normal traffic (False Positive) is not affected by time. 
    \begin{itemize} 
        \item The number of false positives occurring in time window $i$ is $C_{FP}(i)$. 
        They define, ``A False Positive is accounted when a Normal IP address is detected as Botnet at least once during the comparison time frame.''
        \item The time-dependent quantity representing false positives  in time window $i$ is 
        $$FP(i) = \frac{C_{FP}(i)}{N_n(i)}.$$
    \end{itemize}
\end{itemize}
Finally, the time-dependent accuracy metrics, e.g., true positive rate, precision, F1 score, accuracy, etc. are defined as usual using the $TN(i), FN(i), TP(i), FP(i)$ as defined above. 



To summarize, detecting or failing to detect real botnet traffic (or other attack traffic) is time-sensitive, while for normal traffic it is not.
Also, any number of alerts for flows that are related to the same address
should be aggregated into one item for evaluation purposes, 
since the analyst is primarily concerned with the machine-level,
and not directly concerned with the flow-level.
Additionally, these authors define new time-based measures of  
FPR, TPR, TNR, FNR, Presicion, Accuracy, Error rate, and F1 score.
They provide a public tool to calculate these new scores \cite{garcia2014empirical}\footnote{\url{http://downloads.sourceforge.net/project/botnetdetectorscomparer/BotnetDetectorsComparer-0.9.tgz}}.
See Section \ref{sec:virtue}, where we compare these scores to others in a simulation of seven detectors. 

\subsubsection{Common Problems with ML Approaches}
Sommer \& Paxson~\cite{sommer2010outside} review the difficulties in using machine learning in intrusion detection,
and help to explain why it has been less successful when compared with other domains, such as optical character recognition (OCR).
Some of the main issues they highlight involve the lack of quality training data, specifically insufficient quantity of data, training with one-class datasets, and non-representative data causing significant problems.
Additionally they re-emphasize some important practical issues, 
such as the relatively high costs of both false positives and false negatives,
and the ``semantic gap'' referring to the difficulty in interpreting alerts.
All of these factors result in difficulties in performing evaluations,
with the simple statistical metrics of FP \& FN rates being insufficient,
and real-world usability being more important, but more difficult to measure.
The authors emphasize that designing and performing the evaluation is generally
``more difficult than building the detector itself.''


A common problem for  intrusion detection metrics is the base-rate fallacy; e.g., see Axelsson~\cite{axelsson2000base}.
Concisely, the base-rate fallacy is the presence of both a low false positive rate (percentage of negatives that are misclassified) but a high false alert rate (percentage of alerts are false positives, equivalently, $1 -$ precision). 
The base-rate fallacy is caused by high class imbalance, usually orders of magnitude more negatives (normal data) than positives (attack data). 
That is, the denominator of the false positive rate calculation is usually an enormous number; hence, for nearly any detector  the false positive rate can be exceptionally low.  
This can give a false sense of success and it means that ROC curves are only in effect depicting the true positive rate.  
On the other hand, the false alert rate, or simply quantity of false alerts are often important to take into account.

Their overall conclusion is that the intrusion detection problem is fundamentally harder
than many other machine learning problem domains, and that while these techniques are still promising,
they must be applied carefully and appropriately to avoid these additional difficulties.

\subsubsection{Evaluating Other Tools}
Most of these works discussed above focus on IDSs specifically, 
but these difficulties also apply to IPSs, malware detection, and related problems.  
Similarly, these apply regardless of the data source or architecture being considered\textemdash host-based, network-based, virtual machine hypervisor-based, or other approaches.

Additionally, when evaluating other security-related systems, such as firewalls, SIEMs (security information and event management systems), ticketing systems, etc.,
we encounter even more difficulty.
Not only are there no widely accepted datasets, the relevant metrics and the testing methodology are often not considered systematically.
Some of these factors, such as the user experience, integration with current workflow and current tools, etc. 
are inherently harder to quantify, and are often organization-dependent. 
The situation at present is somewhat understandable,
but we maintain that a unified approach to evaluation should help in addressing these areas as well.

%% file: 24-competitions.tex
\subsection{Evaluation criteria for cyber competitions} 
\label{sec:competitions}
Red team \& capture-the-flag (CTF) competitions exercise both offensive and defensive computing capabilities. 
These activities are commonly used as educational opportunities and for organizational self assessment \cite{red-team-article, doupe2011hit, patriciu2009guide, reed2013instrumenting, werther2011experiences, mullins2007cyber}\footnote{Also see DOE's cyber-physical defense competition \url{https://cyberdefense.anl.gov/about/}, DefCon's CTF \url{https://www.defcon.org/html/links/dc-ctf.html},  NCX (NSA's) \url{https://www.nsa.gov/what-we-do/cybersecurity/ncx/}, National Collegiate Cyber Defense Competition (NCCDC) (\url{http://www.nationalccdc.org/}),  and Mitre's CTF (\url{http://mitrecyberacademy.org/competitions/})  among others.}. 
These competitions require a set of resources to be attacked and/or defended and an evaluation criteria to determine winning teams (among other necessities).

In addition to traditional statistical evaluation metrics, the competitions integrate measures of operational viability, such as, the duration or number of resources that  remained confidential, unaltered (integrity measure), or  functional (availability measure), in addition to statistical measures, e.g., true positive rate, etc. 
For example, Patriciu \& Furtuna~\cite{patriciu2009guide} list the following scoring measures for cyber competitions
(for attackers:) the count of successful attacks, accesses to target system, and number of successfully identified open services compared to the total number, an analogue true positive rate,  
(for defenders:) true positive rates for detection (identification) and forensics (classification), time duration to recover from an attack, and downtime of services.

While there is wide variety across competitions, the main trend in evaluation is to augment the usual detection accuracy metrics with some measure of how well an operation remained healthy and unaffected. 
This greatly increases trust in the evaluation procedure because the effect of the security measures on the operational objective are built into the metrics. 
We note that the object under evaluation is usually the participants' skill level, and that significant effort is needed to assemble the test environments.  
While cyber exercise publications often focus on a combination of pedegogy, design, implementation, etc., we only survey the scoring procedures. 

Parallel to such competitions are red-team events and penetration testing as used by security operation centers and software development companies to assess weaknesses in security posture and vulnerabilities in code. 
Penetration testing and red-team testing occupies its own space in the literature, e.g., see \cite{bishop2007penetration, singh2016penetration, randhawa2018mission, stefinko2016manual}, and automated penetration testing tools are becoming prevalent in the commercial market, e.g., \url{www.verodin.com/}   \url{www.attackiq.com/}. 
These works/technologies target an enumeration of weaknesses, not a quantifiable measure for head-to-head comparison---although they can certainly feed such an evaluation! 
We note that DARPA has centered CTFs on the topic  of automated vulnerability analysis and patching, necessitating a scoring metric for this topic (see \href{https://archive.darpa.mil/CyberGrandChallenge/}{\texttt{https://\\
archive.darpa.mil/CyberGrandChallenge/}}). 
Below we discuss a few works that give novel evaluation metrics for cyber competitions. 

\subsubsection{iCTF's Attacker Evaluation} 
Doup\'{e} et al.~\cite{doupe2011hit} describes the 2010 International Capture the Flag Competition (iCTF), which employed a novel ``Effectiveness'' score for each attacker.  
For each service, $s$, and time $t$ the binary functions $C, D, OT$, taking values in $[0,1]$, are defined as follows: 
$C(s,t)$ is a binary function that indicates criticality of service $s$ at that time $t$; specifically, it indicates if the function is in use for this application. 
$D(s,t)$ encodes risk to an attacker, e.g., being detected, and in this competition was simply the opposite bit as $C$, punishing attacks on unused services. 
$OT(s,t) $ is the indicator function for when $C-D$ is positive and represents the ``Optimal Attacker''. 
For an attacker $a$, $A(s,t,a)\in [0,1]$ represents the risk to service $s$ by attacker $a$ at time $t$.  
Toxicity is defined as 
$$T(a,s) = \int_t A(a,s,t)(C-D)(s,t) dt$$
a score that is increasing with the effectiveness of the attacker.  
Note that toxicity ($T$) is maximal when $A = OT$.  
Hence, the final score is the normalized toxicity, 
$T(a,s)/ Z$ with $Z(s): = \int OT(C-D)(s,t)dt.$

\subsubsection{MIT-LL CTF's CIA Score}
Werther et al.~\cite{werther2011experiences} describes the MIT Lincoln Laboratory CTF exercise, in which teams are tasked with protecting a server while compromising others' servers. 
A team's defensive score is computed as 
$$D:= \sum_{x\in\{c,i,a\}} W_x  x,$$ 
where 
\begin{enumerate}[leftmargin = *] 
\item $c$ is the percent of the team's flags not captured by other teams (confidentiality), 
\item $i$ is the percent of the team's flags remaining unmodified (integrity), and 
\item $a$ is the percent of successful functionality tests (availability). 
\end{enumerate}
Weights $W_{\{c,i,a\}}$ allow flexibility in this score. 
The offensive score is $O:=$ fraction of flags captured from other teams' servers, and the total score is 
$$W_d D + (1-W_d) O,$$ 
with   parameter $W_d\in (0,1)$  encoding the tradeoff between offensive and defensive scores. 
This is an appealing metric because it intuitively captures all three facets of information security\textemdash confidentiality, integrity, availability. 

\subsubsection{DARPA's Vulnerability Analysis \& Patching Grand Challenge CTF Score} 
DARPA's Grand Challenge (\href{https://archive.darpa.mil/CyberGrandChallenge/}{\texttt{archive.darpa.mil/\\CyberGrandChallenge/}}) focused on vulnerability analysis and patching. Each team's entry, a fully autonomous ``cyber reasoning system'' (CRS) was connected to the referee server that served potentially vulnerable services (binaries) simultaneously to each team's CRS. 
Teams then had the ability to discover vulnerabilities in these services and patch their instances, send ``challenges'', that is exploit code to prove a vulnerability exists in opponent's binaries, and build network IDS rules to protect their vulnerable binaries from others' exploits. 
The competition is comprised of a sequence of rounds, and scoring is administered for each CRS per service per round and then summed. 
A CRS's service score in a given round is computed as the product of \textit{availability}, \textit{security}, and \textit{evaluation} subscores.  
To achieve the \textit{availability} subscore of a CRS's service for a round, the referee would test the service for proper functionality, which may be compromised by an opponents exploit, and improper patch, or when the service is down because it is being patched. 
This score lies in the range 0-100. 
The \textit{security} subscore of a CRS's service in a round is 2 unless the service was exploited by an opponent, then it was decremented in that round to 1. 
Finally, the \textit{evaluation} subscore of a CRS's service,
$1 + \frac{x}{N}$, where $N = 6$ was the number of opponents, and $x$ is the number of successful ``chalenge'' exploits of that service in this round. 
 \cite{avgerinos2018mayhem}
 Also see \href{ http://webcache.googleusercontent.com/search?q=cache:r9zSUTMtXhIJ:www.phrack.org/papers/cyber_grand_shellphish.html&hl=en&gl=us&strip=1&vwsrc=0}{\texttt{www.phrack.org/papers/\\cyber\_grand\_shellphish.html}}.  
Hence, those that could quickly identify and prove (exploit) vulnerabilities as well as  patch their own software with little down time were most rewarded. 

We conclude with a major lesson for proposing quantifiable evaluation frameworks that is illustrated well by the DARPA challenge scoring---that the method used to reduce security to a single number can promote undesirable security postures. 
For example, see a DARPA challenge  competitor's web page\footnote{\url{http://webcache.googleusercontent.com/search?q=cache:r9zSUTMtXhIJ:www.phrack.org/papers/cyber_grand_shellphish.html&hl=en&gl=us&strip=1&vwsrc=0}} where they claim to have simulated an entry that does nothing---that is, the entry takes no actions to find, patch, exploit or defend vulnerabilities---and concluded that this null entry would have placed third in the competition. 
While this may be a desired result---e.g., perhaps the current state of automated patching affects availability so much that it is not yet viable---it is a firm reminder that quantitative evaluations are vulnerable to exploit themselves! 
In Section \ref{sec:virtue} we provide an example of our model configured for a cyber competition, and we simulate a variety of strategies to test its efficacy and ability to be `gamed' by an unrealistic security strategy.

%% file: 26-cost-benefit.tex
\subsection{Cost-benefit analyses of security measures}
\label{sec:cost_benefit}
There is a robust literature that bloomed around 2005 providing quantifiable cost-benefit analysis of security measures using applied economics. 
Arising from the tension between the operational need for security and the organization's  budget constraints, these researches provide frameworks for quantifiable comparison of security measures. 
The clear goal of each work is assisting decision makers (e.g., C-level officers) in optimizing the security-versus-cost balance. 
To quote Leversage \& Byres~\cite{leversage2008estimating}, 
\begin{quote}
    One of the challenges network security professionals face is providing a simple yet meaningful estimate of a system or network's security preparedness to management, who typically aren't security professionals.
\end{quote}

This goal is also emphasized in other, more general works on security metrics, such as Andrew Jaquith's book on security metrics\cite{jaquith2007security} and NIST Special Publication 800-55\cite{chew2008nist}.  These both describe how to choose or create suitable metrics for an organization, and how to relate these to the organization's mission.
These works focus on metrics which are easily measured and easily understandable, such as the proportion of machines with a specific vulnerability, and then aggregating these into higher-level metrics, such as the proportion of machines which are non-compliant with policy.  
What differentiates the following works is the focus on economic metrics and estimates which should be applicable to any organization.

While many different models exist, overarching trends are to enumerate/estimate (a) internal resources and their values, (b) adversarial actions (attacks) and their likelihood, and (c) security measures'  costs and effects, then use a given model to produce a comparable metric for all combinations of security measures in consideration. 
This subject bleeds from academic literature into advisory reports from government agencies and companies \cite{ansi-report, ponemon-resilience},   textbooks for management \cite{gordon2006managing, tipton2007information}, 
and security incident summary costs and statistic reports  \cite{fbi-report, ponemon-cost, fireeye-cost}.

The main drawback is all proposed models rely on untenable inputs (e.g., likelihood of a certain attack with and without a security tool in place) that are invariably estimated and often impossible to validate. 
Academic authors are generally open about this as are we.  
Perhaps surprisingly, our survey of the literature did not identify use of sensitivity analysis to identify the most critical assumptions, a reasonable step to identify which inputs are most influential, especially when validation of input assumptions is not possible. 
In response, for our model we provide such a discussion in Section \ref{sec:full-model}.  

A prevalent, but less consequential drawback is a tendency to oversimplify for the sake of quantification. 
This often results from unprincipled conversions of incomparable metrics (e.g., reputation to lost revenue), or requiring users to rank importance of incomparable things. 
The outcome is a single quantity that is simple to compare but hard to interpret. 

User studies shows that circa 2006, many large organizations used such models as anecdotal evidence to support intuitions on security decision \cite{rowe2006private}. 
The advantages are pragmatic\textemdash these models leverage the knowledge of security experts and external security reports to (1) reason about what combination of security measures is the ``best bang for the buck'', and (2) they provide a financial justification required by chief financial officers to move forward with security expenditures \cite{ansi-report}.

\subsubsection{SAEM: Security Attribute Evaluation Method}
Perhaps the earliest publication on cost-benefit analysis for information security, Butler~\cite{butler2002security} provides a detailed framework for estimating a threat index, which is a single value representing the many various expected consequences of an attack. 
Working with an actual company, Butler describes examples of the many estimates in the workflow. 
Users are to list (1) all threats, e.g,. 28 attacks were enumerated by the company using this framework each in three strengths, (2) all potential consequences with corresponding metrics, e.g., loss of revenue measured in dollars, damaged reputation measured on a 0-6 scale, etc., and (3) the impact of each attack on each consequence. 
Weights are assigned to translate the various cost scales into a uniform ``threat index'' metric; note that this step allows a single number to represent all consequences, but is hard to interpret.  
Next, the likelihood of each attack is estimated, and the weighted average gives the threat index per attack. 
The per attack threat indices are summed to a single, albeit hard-to-interpret number. 
By estimating the effect of a desired security measure on the inputs to the model, analysts can see the plot of costs for each solution versus the change in threat index. 
Notably, authors mention that uniformly optimistic or pessimistic estimates will not change rankings of solutions, and suggest a sensitivity analysis, although none is performed.

\subsubsection{ROSI: Return on Security Investment} 
Sonnenreich et al. \cite{sonnenreich2006return} and Davis ~\cite{davis2005return} discuss a framework for estimating the Return on Security Investment (ROSI). 
The calculation requires estimation of the Annual Loss Expected ($ALE$). 
Tsiakis et al. \cite{tsiakis2005economic} provide three formulas for estimating $ALE$. 
One example is to let $O_i$ be the set of attacks, $I(O_i)$ the cost of the attacks, $F(O_i)$ the frequency of the attack, and then $ALE = \sum I(O_i)F(O_i)$.  
ROSI is a formula to compute the percent of security costs saved if implemented. 
It requires users to estimate the percent of risk mitigated $M$ by the security measure and the cost of measure $C$. 
Then the expected costs are $ALE * M - C$, and ROSI (the percent of cost returned) is $(ALE*M-C)/C$. 
These authors expect estimate formulas to vary per organization, point to public cost-of-security reports, e.g., \cite{fbi-report} to assist estimation, and suggest internal surveys to estimate parameters needed. 
They go on to say that ``accuracy of the incident cost isn't as important as a consistent methodology for calculating and reporting the cost'', a dubious claim.  

\subsubsection{ISRAM: Information Security Risk Analysis Method}
Karabacak~\cite{karabacak2005isram} introduces ISRAM, a survey procedure for estimating attack likelihood and cost, the two inputs of an $ALE$ estimate. 
For both attack likelihood and attack costs, a survey is proposed. 
Each survey question (producing an answer which is a probability) is given a weight, and the weighted average in converted to a threat index score, which is averaged across participants. 
The ALE score is the product of these two averages. 

\subsubsection{Gordon-Loeb Model} 
\label{sec:gordon-loeb}
Perhaps the most influential model is that of Gordon and Loeb (GL Model), which provides a principled mathematical bound on the maximum a company should spend on security in terms of their estimated loss.  
See the 2002 paper~\citep{gordon2002economics} for the original model. 
Work of Gordon et al.~\cite{gordon2015externalities} extends the model to include external losses of consumers and other firms (along with costs only to the private firm being modeled).

To formulate the GL model, let $L$ denote the monetary value of loss from a potential cyber incident, $v$ the likelihood of that incident, and $s(z)~:~[0,~\infty)~\to~(0,v]$ denote the likelihood of an attack given $z$ dollars are spent on security measures. 
Initial assumptions on $s$  are that $s(0) = v > 0$, $s$ is twice differentiable, and $s$ strictly convex; e.g., $s(z)= v\exp(-az)$ for $a>0$ is a particular example.  
It follows that $vL$ is their $ALE$ estimate. 
The goal is to optimize the expected cost, $s(z) L - z$ for $z$ positive. In the initial work Gordon \& Loeb show that for two classes of $s$ satisfying the above assumptions, $z*: = $argmin$_z (s(z)L - z),$  the spending amount that minimizes expected costs, satisfies 
\begin{equation}
\label{gl-rule}
    z* \leq L/e.
\end{equation} 
That is, optimal security will cost no more than the $1/e\approx 37\%$ the expected loss of the attack \cite{gordon2002economics}!

Follow-on mathematical work has shown this bound to be sharp and valid for a much wider class of functions $s$ ~\cite{lelarge2012coordination, baryshnikov2012security}. Specifically, the work of Baryshnikov~\citep{baryshnikov2012security} is particularly elegant with mathematical results so striking they are worth a summary. 
Let $X$ be the set of all security actions a firm could enact, $Z(A)$ the cost of a set of actions $A\subseteq X$, and $S(A)$ the likelihood of an attack after actions $A$ are enacted. 
Baryshinkov assumes enactable collections of actions are measurable, and $Z$ is a measure; this is a mild assumption and its real-world meaning is simply that the cost of disjoint collections of security actions will be additive, i.e., $Z(A_1\dot\cup A_2 ) = Z(A_1)+Z(A_2)$.
Next, $S$ is also a set function with  $S(A)$  interpreted as the likelihood of an attack after actions $A$ are enacted. 
There are two critical assumptions\textemdash 
\begin{enumerate}[leftmargin = *] 
\item  $S(A_1\dot\cup A_2 ) = S(A_1)S(A_2)$, so indeed $u:= -\log S$ is a measure.
\item $U$ is a non-atomic measure, i.e., any $A$ can be broken into smaller $U$-measurable sets.
\end{enumerate}
These assumptions are made to satisfy the hypotheses of Lyapunov's convexity theorem (see \cite{tardella1990new, liapounoff1940fonctions}). 
Finally,  set 
$$s(z_0) = \inf_{Z(A)\leq z_0} S(A),$$
the likelihood of an attack given one has enacted the optimal set of security actions that cost less than $z_0$.  
Lyapunov's theorem furnishes that the range of vector-valued measure $(Z,U)$ is closed and convex.  
The closedness, implies that for any $z_0$ (amount of money spent), the optimal set of counter measures exists,  while  the convexity can be used to show that the $z*$ (the optimal cost) satisfies the 37\% rule (Equation~\ref{gl-rule})!

This dizzying sequence of mathematics is striking because it starts with few and seemingly reasonable assumptions and proves the cost of optimal security is bounded by $37\%$ of potential losses. 
The conundrum of these results is they are deduced with no real-world knowledge of a particular organizations, security actions, costs, or attacks. 
While the assumptions seem mathematically reasonable, e.g. ``$s$ is convex'' translates to ``decreasing returns on investment (the first dollar spent yields more protection than the next)'', the result, the 37\% rule,  presupposes the solution to a critical question\textemdash  \textit{that for any given dollar amount, $z$, the optimal security measures with cost less than $z$ will be found}. 
No method for finding an optimal set of measures is given or widely accepted. 

Gordon et al.~\citep{gordon2016investing} focuses on ``insights for utilizing the GL model in a practical setting''. 
Since the model is formulated as optimizing a differentiable function, the optimum occurs when $s' = 1$, or equivalently, the increment of spending in which the marginal likelihood of attack is estimated at 1 is the amount to spend. 
The authors work with a company as an example, and the company is tasked with identifying resources to protect,  the losses if each is breached, and change in likelihood for each \$1M spent. 
In practice this model mimics the many other works in the area. 
The burden is on the company to  estimate cost, likelihood, and efficacy of potential attacks and countermeasures, and then the reasoning is straightforward. 
On the other hand, the 37\% rule gives an indicator if an organizations' security expenses are non-optimal. 
See Section \ref{sec:virtue} for an application.

\subsubsection{Leversage \& Byres' Mean Time to Compromise Estimate} 
Research by Leversage \& Byres \cite{leversage2008estimating} uses the analogy of burglary ratings of safes, which is given in terms of time needed for one to physically break into the safe, as a way to quantify security. 
Specifically, the research seeks an estimate of the average time to compromise system. 
Network assets are divided into zones of protection levels and network connectedness is used to create an attack graph using some simplifying assumptions, e.g., a target device cannot be compromised from outside  its zone. 
Attackers are classified into three skill levels, and functions are estimated that produce the time to compromise assets given the attacker's level and other needed estimates, such as, average number of vulnerabilities per zone. 
Finally, a mean time to compromise can be estimated for each adversary level using the paths in the attack graph to targets and estimated time functions. 
While this model still requires critical inputs that lack validated methods to estimate, the work addresses the problem of quantifying security in a different light. 
Unlike the other models discussed here, it embodies the fact that time is an extremely important aspect of security for two reasons: (1) The more adversarial resources are needed to successfully compromise a resource, the less likely they are to pursue/succeed; (2)  The more time and actions needed between initial compromise and target compromise, the more chance of detection and prevention before the target is breached \cite{ponemon-cost}.

\subsubsection{Other works on quantifying security} 
Tangential to the three research areas discussed above are various researches and non-academic reports that address quantifiability of security. 

Vendor and government reports are common resources for estimating costs based on historical evidence. 
Broad statistics about the cost and prevalence of security breaches are provided annually by the US Federal Bureau of Investigation (FBI)~\cite{fbi-report}. 
More useful for estimating costs of a breach are industry reports that provide statistics conditioned on location, time, etc.~\cite{ansi-report, ponemon-resilience, ponemon-cost, ponemon-cost-2018, fireeye-cost}. 
Notably, Ponemon's Cyber Cost report gives the average monetary cost per record compromised per country per year\textemdash \$225 \& \$233 per record in the US in 2017, 2018 respectively they report\textemdash an essential estimate for all economic models above. 
Further, Ponemon's reports that if the mean time to compromise (MTTC) was under 30 days, the average increase total cost was nearly \$1M less than breaches with MTTC greater than 30 days. 

Acquisiti et al.~\cite{acquisti2006there} seek the cost of privacy breaches through statistical analysis of the stock prices of many firms in the time window surrounding a breach. 
Their conclusion is short-term negative effects are statistically significant, but longer term are not.

See Rowe \& Gallaher~\cite{rowe2006private} for results of a series of interviews with organizations on how security investments decisions are made (circa 2006). 
Anderson \& Moore~\cite{anderson2006economics}  provide a 2006 panoramic review of the diverse trends and disciplines influencing information security economics. 

Verendel~\cite{verendel2009quantified} provides a very extensive pre-2008 survey of researches seeking to quantify security, concluding that ``quantified security is a weak hypothesis''. 
That is to say, the methods proposed lack repeated testing resulting in refinement of hypotheses and ultimately validation through corroboration. 

%% file: 30-proposed-approach.tex
\section{New Evaluation Framework}
\label{sec:new_approach}
Our goal is to provide a comprehensive framework for accruing security costs that can be flexible enough to accommodate most if not all use cases by modeling and estimating costs of defensive and offensive measures modularly. 
In particular, we target the use case of comparing tools, procedures, or operators, e.g. and especially, for security operation centers and cyber competitions. 
By design the model balances the accuracy of detection and prevention capabilities, the resources required (hardware, software, and human), and the timeliness of detection and incident response. 
Viewed alternatively, the model permits cost estimates for true negative (not under attack), true positive (triage and response costs), false negative (under attack without action), and false positive (unnecessary investigative) states. 


Our approach can be seen as adopting the same general cost-benefit framework as the works in Section \ref{sec:cost_benefit},
and incorporating the more specific metrics described in Sections \ref{sec:ids} and \ref{sec:competitions} 
to address the other two use cases,
namely IDS evaluation and competition events.
More specifically, to evaluate the impact of any technology, policy, or practice, we estimate the change in the total cost (\(C_{\text{Total}}\))
by estimating the costs of breaches ($\sum C_{\text{Breach}_i}$)
and the cost of all network defenses ($C_{\text{Defense}}$). 
The costs of network defense ($C_{\text{Defense}}$)
can be considered a combination of labor costs $C_{\text{L}}$ and resource costs $C_{\text{R}}$. 
\begin{equation}
\label{eqn:total-costs}
    C_{\text{Total}} = \sum C_{\text{Breach}_i} + C_{\text{Defense}}.
\end{equation}
The attack cost, 
$\sum C_{\text{Breach}_i}$ is analogous to the Annual Loss Expected ($ALE$) following Section \ref{sec:cost_benefit}, with the difference being that $\sum C_{\text{Breach}_i}$ covers an arbitrary time period, and can include actual or estimated losses.
Note that these breach costs include both direct costs (monetary or intellectual property losses) as well as less direct losses such as reputation loss, legal costs, etc.
Defense costs $C_{\text{Defense}}$ include all costs of installing,
configuring, running, and using all security mechanisms and policies.
While effective defenses will reduce the number of breaches expected,
effective incident response will reduce the impact (and therefore the cost) of any specific breach $C_{\text{Breach}_i}$,
so both approaches would be expected to reduce $\sum C_{\text{Breach}_i}$,
at the cost of somewhat increasing $C_{\text{Defense}}$.
The defense cost includes both resource costs and labor costs.
Both of these will generally include up-front as well as ongoing costs.
Ongoing costs can vary over time, and can depend on adversary actions,
because analysts will be reacting to adversary actions when detected.

When comparing security tools and/or  procedures, we strive to incorporate the total costs of all candidate systems,
meaning the defense costs $C_{\text{Defense}}$
plus the projected breach costs $\sum C_{\text{Breach}_i}$ above.
A typical analysis might compare a baseline of no defenses,
(meaning $C_{\text{Defense}} = 0$  and maximal $\sum C_{\text{Breach}_i}$,)
versus current practices, versus new proposed system(s).
The total costs ($C_{\text{Total}}$) will be positive in all cases,
but successful approaches will minimize this total.

For a simple example, consider an enterprise
implementing a policy that all on-network computers
must have a particular host-based anti-virus alerting and blocking system. 
Such a change will incur an upfront licensing fee, costs of hardware needed to store and process alerts, 
labor costs for the time spent installing and configuring, time spent responding to alerts, 
and a constant accrual of costs in terms of memory, CPU, and HD use per host per hour.
However, these $C_{\text{Defense}}$ costs will presumably be offset
by a reduced $\sum C_{\text{Breach}_i}$.
Estimating all of these costs included in $C_{\text{Defense}}$ is relatively simple,
however estimating $\sum C_{\text{Breach}_i}$ is more difficult.

We rely on many of the same cost estimates as the works in Section \ref{sec:cost_benefit},
which does present some practical difficulties,
especially when estimating probability or costs of an attack. 
 While a clear drawback, these difficulties in estimation are unavoidable.
We make two concessions:
(1) First,  this is often not a hindrance when using the model for comparison of similar tools/procedures. 
When populating inputs with estimates (e.g., attack cost or host resource costs), inaccurate input values may indeed invalidate the accuracy of the output value (that is, the actual overall cost may be incorrect), but when comparing similar tools/procedures, the ranking provided by our cost model will be accurate even if the exact estimated costs are not (provided the input cost estimates are held constant across the instantiations of the model). 
Furthermore, we provide estimates for costs based on research  to be used as defaults in Section~\ref{sec:quantifying_costs}. 
(2) Second, we provide a sensitivity analysis in Section \ref{sec:full-model} that illuminates the affect of each parameter. This provides guidance to the user on what estimates should be most accurate, or, if a user cannot accurately estimate influential parameters, they  can at least vary these parameters within an acceptable range to gauge results.


The main benefit of the approach is the flexibility. 
This approach can be applied to a wide variety of technology, procedural, or policy changes and compare them head-to-head. 
For example, one may wish to compare a new SIEM, which bears large initial costs in terms of licensing, configuration, and training but enhances efficiency of operators, to a new procedure for handing off incidents between operators to increase efficiency, 
to a new IDS that will increase accuracy of alerts. 
Admittedly, in our examples we are primarily considering the case of IDS evaluation. 
(Specific examples for using the model for such evaluations are the topic of Section \ref{sec:examples}.)

This section defines and itemizes attack and defense costs in Subsections \ref{sec:defining_attacks_and_breaches} \& \ref{sec:defense-cost}. 
We strive for relatively fine-grained treatment of costs (e.g., breaking attack cost models into kill-chain phases), permitting one to drill down into costs if their data/estimates permit detailed analysis, or to stay at a more general level and model with coarser granularity. 
Subsection \ref{sec:full-model} describes how these are combined, and
this section concludes with our estimates for quantifying the main components in Subsection \ref{sec:quantifying_costs}.

\input{31-define-attacks.tex}

\input{311-attack-costs.tex}

\input{32-defense-costs.tex}
\input{321-labor.tex}

\input{322-resource.tex}
\input{33-full-model.tex}

\input{34-quantifying-costs.tex}

%% file: 31-define-attacks.tex
\subsection{Attacks and Breaches: Definition and Cost Model}
\label{sec:defining_attacks_and_breaches}

We define a ``breach'' as any successful action by an attacker that compromises any of the familiar triad of confidentiality, integrity, or availability. 
In our examples, we primarily consider the costs of losing confidentiality and integrity, but losses of availability also have measurable and potentially significant costs.  The relative importance of these potential costs will be highly dependent on the organization.



Building on this, we consider an ``attack'' to be a series of actions which,
if successful, will lead to a loss or corruption of data or resources. 
The attack begins with the first actions that could lead to this loss, and the attack ends when these are no longer threatened.
For example, an attacker may re-try a failed action several times before adapting or giving up, and this would all be considered part of the same attack.
An attack can potentially be thwarted by both automated tools and manual response of the SOC.

If each attack were instead viewed as one atomic event,
this type of reaction by network defenders would not be possible within that framework;
however, real attacks almost always involve a sequence of potentially-detectable attacker actions.
For example, the ``cyber kill chain'' model \cite{hutchins2011intelligence} describes a seven-phase model of the attacker's process,
beginning with reconnaissance, continuing through exploitation, command and control, and ending with the attacker completing whatever final objectives they may have.
At that point, the breach is successful.
As the authors describe:
\begin{quote}
The essence of an intrusion is that the aggressor must develop a payload to breach a trusted boundary,
establish a presence inside a trusted environment,
and from that presence, take actions towards their objectives,
be they moving laterally inside the environment or violating the confidentiality,
integrity, or availability of a system in the environment.
The intrusion kill chain is defined as reconnaissance,
weaponization, delivery, exploitation, installation, command and control (C2),
and actions on objectives.
\end{quote}
The authors later describe how this model can map specific countermeasures
to each of these steps taken by an adversary,
and how this model can be used to aid in other areas such as forensics and attribution.

Other authors have expanded this kill chain model
to related domains such as cyber-physical systems \cite{hahn2015multi}
or proposing related approaches based on the same insights \cite{caltagirone2013diamond}.
The creators of the STIX model discuss this kill-chain approach in some depth
when presenting their STIX knowledge representation \cite{barnum2012standardizing}.
They define a ``campaign'' as
``a set of attacks over a period of time against a specific set of targets to achieve some objective.''

Our definitions of ``attack'' and ``breach'', discussed above,
is a simplified view of these same patterns.
In this case, the specific sequence of actions is less important than the general pattern:
an ``attack'' consists of a series of observable events,
which potentially leads to a ``breach'' if successful.
The events that make up an attack can be grouped into ``phases'',
where one phase consists of similar events,
and ends when the attacker succeeds in progressing towards their objectives.
An example would be dividing the attack into seven phases corresponding to the kill chain above;
however, we generally make no assumptions about what may be involved in each phase, or how much time may occur between each phase,
only that they occur sequentially,
and that succeeding in one phase is a prerequisite for the next.
Also note that because of how ``attack'' and ``breach'' are defined, the user has the option to combine or separate related attacks in whatever way is most intuitive or convenient, for example modeling an APT campaign either as one long attack with one overall objective, or else a series of more conventional shorter attacks each with more limited objectives.  Either approach should produce equivalent results.

%% file: 311-attack-costs.tex
\subsubsection{Breach Costs Model, $C_{\text{Breach}}$}
\label{sec:breach_costs}
The general pattern for the attack is that each phase incurs a higher cost than the previous phase,
until the maximum cost is reached when the attacker succeeds.
Within each phase, the cost begins at some initial value,
then increases over time until it reaches some maximum value for that phase.
Consider an attacker with user-level access to some compromised host.
Initially, the attacker may make quick progress in establishing one or more forms of persistence,
gathering information on that compromised system, evaluating what data it contains, etc.
However, over time the attacker will make maximal use of that system,
and will need to move on to some other phase to continue towards their objectives.

\begin{figure}[ht]
\centering
\includegraphics[width=.75\linewidth]{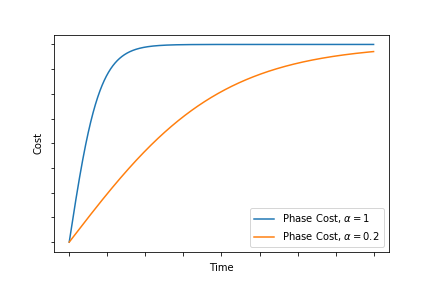}
\caption{Plot of a proposed cost $C_{\text{phase}}(t; a,b,\alpha): =  b/(1+c \exp(-\alpha t)) $ versus time $t$ for one phase of an attack, where $c = b/a - 1$ with  $a$ the starting cost at $t=0$, $b$ the maximal cost (limit in time), and term $\alpha$ determines how quickly this maximum cost is approached. }
\label{fig:cost-vs-time-phase}
\end{figure}

To model the phase of an attack we simply require a cost function that increases over time and asymptotically approaches the maximum cost; at it's most basic, this can simply be a step function indicating the full cost was incurred at the moment of the attack. 
For a more flexible model, the cost versus time for any phase of the attack can be seen in Figure \ref{fig:cost-vs-time-phase},
as represented by the equation $C_{\text{phase}}(t; a,b,\alpha): =  b/(1+c \exp(-\alpha t))$,
where $c = b/a - 1$ with  $a$ the starting cost at $t=0$, $b$ the maximal cost (limit in time), and term $\alpha$ determines how quickly this maximum cost is approached.
In the worst case, where $\alpha$ is relatively large, this cost versus time curve is approximately a step function.

\begin{figure}[ht]
\centering
\includegraphics[width=.75\linewidth]{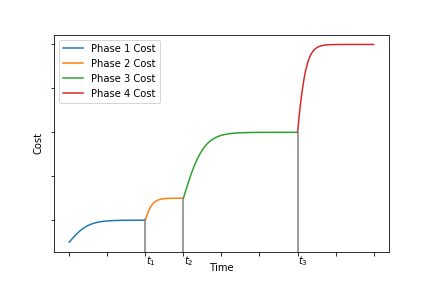}
\caption{Plot of cost versus time for a proposed model of all phases of an attack, specifically, $C_{\text{Breach}}(t) = \sum C_{\text{phase}_i}(t-t_i)$ with phase $i$ beginning at time $t_{i}$. }
\label{fig:cost-vs-time-attack}
\end{figure}

Because an attack is composed of several of these phases,
if we assume that each phase is more severe and more costly than the previous,
we can view the cost versus time for the attack overall as seen in Figure \ref{fig:cost-vs-time-attack}.
This can be represented by a sum of the cost of all phases, which using the equation above would be 
$C_{\text{Breach}}(t) = \sum C_{\text{phase}_i}(t-t_i)$ 
with phase $i$ beginning at time $t_i$. 
As $t$ increases, this will approach the maximum cost for this breach. 

\begin{figure}[ht]
\centering
\includegraphics[width=.7\linewidth]{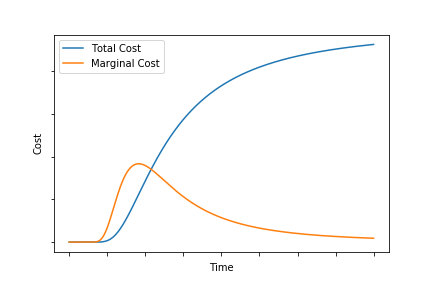}
\caption{Plot of an proposed S-curve estimate of attack cost $C_{\text{Breach}}(t) = b \exp(-\alpha / t^2)$ and corresponding marginal cost (derivative) $2\alpha b \exp(-\alpha/t^2)/t^3$ ignoring individual phases of an attack. }
\label{fig:s-curve}
\end{figure}

This model is crafted to give flexibility based on the situation. 
In cases where we have sufficient data on a real attack, this approximation may be unneeded, and one can replace the curves above with observed costs for each phase. 
On the other hand, when estimating cost of a general attack without specific cost versus time data, we propose two options. 
First, one may consider estimating each phase's cost as constant and estimate the attack as a series of steps (effectively letting each phase's $\alpha\to \infty$ in the proposed cost models above).  
Secondly, one may ignore the individual phases and approximate the total cost of the attack as an S-curve, such as $b \exp(-\alpha / t^2)$, where $b$ denotes the maximal cost of an attack over time, and $\alpha$ controls how fast the attack cost approaches $b$.  
See Figure~\ref{fig:s-curve}. 

This second approach is useful when the total cost of a breach can be estimated, but the individual phases of an attack either cannot be modeled well or are not the primary concern. 
For example, this approach may be more useful when estimating future breach costs for planning purposes. 
Intuitively, this cost estimation gives a marginal cost of $2\alpha b \exp(-\alpha/t^2)/t^3$, a skewed bell curve. 
The shape of this curve matches the intuitively expected costs for a common attack pattern---beginning with low-severity events such as reconnaissance,
reaching maximum marginal cost as the attack moves laterally,
exfiltrates data, or achieves its main objectives, and then tapers off the accrual of costs. 
That is, over time, after the main objectives have been completed,
and the maximum cost is being approached,
the attack will again reach a lower marginal cost simply because few or no attacker objectives remain.
To view this another way, this model captures the common-sense view that attacks should be stopped as early as possible,
and that stopping an attack after it has largely succeeded provides little value.
Interestingly, modeling a particularly slow-moving attacker, or a particularly fast one, can be achieved by varying $\alpha$. 
In practice one would fit the two parameters to their data/estimates. 
Examples are given in Section \ref{sec:quant-breach-cost} and \ref{sec:virtue}. 











%% file: 32-defense-costs.tex
\subsection{Defense Cost Model}
\label{sec:defense-cost}

As mentioned above, the total costs include both the breach costs, discussed above in \ref{sec:defining_attacks_and_breaches}, as well as the defense costs, $C_{\text{Defense}}$.
This defense cost can be split into labor and resource (e.g., hardware) costs, denoted $C_{\text{L}}$, $C_{\text{R}}$, respectively, 
\begin{equation}
\label{eq:defense-cost}
                     C_{\text{Defense}} = C_{\text{L}}  + C_{\text{R}}.
\end{equation}
Both the labor costs  and resource  costs  can be sub-divided into several terms for easier estimation.
These can represented as a sum of the following:
\begin{itemize}
\item initial costs, $C_{\text{I}}$, covering initial install, configuration, and related tasks,  
\item baseline costs, $C_{\text{B}}$, covering ongoing, normal operation when no alerts are present, 
\item alert triage costs, $C_{\text{T}}$, representing the cost of determining if an alert is a true positive or a false positive, 
\item incident response costs, $C_{\text{IR}}$, representing the costs of responding to a real incident after it is detected and triaged. 
\end{itemize}
This can be summarized as the following equations:
\begin{align} 
\label{eqn:defense-costs}
C_{\text{L}} &= C_{\text{I}_L} + C_{\text{B}_L} + C_{\text{T}_L} + C_{\text{IR}_L}\\
C_{\text{R}} &= C_{\text{I}_R} + C_{\text{B}_R} + C_{\text{T}_R} + C_{\text{IR}_R}
\end{align}










%% file: 321-labor.tex
\subsubsection{Labor Cost Model, $C_L$}
\label{sec:labor_costs}
Labor costs of analyst time and other technical staff time are a significant cost for many organizations.
These costs can be sub-divided as described above,
into initial costs, baseline costs, triage costs, and incident response costs,
which allows the labor costs to be related directly to the sensor behavior and the status of any attacks.  
See Table \ref{tab:labor_costs} with functional models to accompany these descriptions. 

The initial labor costs, $C_{\text{I}_L}$, covers any initial installation, configuration, and all related tasks such as creating/updating any documentation.  
This also includes the costs of any required training for both analysts and end users.  

The baseline labor costs, $C_{\text{B}_L}$, covers normal operation when no alerts are present.  
This would include any patching, routine re-configuration, etc.

The alert triage labor costs, $C_{\text{T}_L}$,
represents the cost of determining if an alert is a true positive or a false positive.  
Note that the time needed to triage any alerts can depend significantly on their interpretability.  
For example, an alert giving ``\texttt{anomalous flow from IP <X>, port <x> to IP <Y>, port <y>}'' would be less useful than
``\texttt{Unusually low entropy for port 22(ssh),\\
this indicates un-encrypted traffic where \\
not expected}''.

The incident response labor costs, $C_{\text{IR}_L}$,
represents the costs of responding to a real incident after it is detected and triaged.
The actual cost of this can vary over a large range,
but we can make many similar observations as in Section \ref{sec:breach_costs}\textemdash
the attack can be considered a series of discrete events,
grouped into phases of escalating severity and cost,
and that each phase reaches some maximum cost before potentially advancing to the next phase.
Overall, we can model the costs of incident response with a sigmoid function, similar to the attack costs model:
$C_{\text{IR}_L}(t) = b \exp(-\alpha / t^2),$ 
with parameters $\alpha, b$ fit to incident response costs data if available. 
Like the attack costs model, if we have data from an actual observed attack, we then no longer need this model, and can calculate this cost directly from available information.

\begin{table}[b!]
\centering
\begin{threeparttable}
   \vspace{-.1cm}
   \centering 
    \caption{Labor Costs}
    {
    \begin{tabular}{l c c }
    \toprule
    Notation (Cost) & Analysts	& End Users \\ 
    \midrule
    $C_{\text{I}_L}$ (Initial) & $c_1$ & $c_2$ \\ 
    $C_{\text{B}_L}$ (Baseline)  & $x_1t$ & $x_2t$ \\ 
    $C_{\text{T}_L}$ (Triage) & $y_1n$ & $0$ \\
    $C_{\text{IR}_L}$ (Incident Response) & $\sum f_i(t) $ & $\sum f_i(t)$ \\ 
    \bottomrule
    \end{tabular}%
    } 
    \label{tab:labor_costs}%

\begin{tablenotes}[para]
\small
  Table of labor costs for analysts and end users. Note that $f_i(t)$ is the cost estimate function, and is not needed if real cost data is available.
\end{tablenotes}
\end{threeparttable}
\end{table} 

If there is any noticeable burden or productivity impact to the end user, this must also be included.
These costs can be categorized in the same way as above.
The initial costs would include the costs of any required setup and training, as mentioned previously. 
The baseline costs would include any possible impact on the end user from normal operation,
such as updating credentials, maintaining two-factor authentication, etc.
Triage costs could in principle apply to both analysts and end users; 
however, generally end users will not be involved in or aware of this process, 
so in most cases this would not contribute to costs.
Incident response may also impact users, for example due to re-imaging machines, or due to network resources being unavailable during the response.
Like above, the impact on users can either be calculated based on real event data,
or estimated using a similar model as the analysts' costs.






%% file: 322-resource.tex
\subsubsection{Resource Cost Model, $C_R$}
\label{sec:resource_costs}
Resource costs are another significant component of overall costs of network defense.
These can be broken down similarly to the labor costs above, into initial costs
$C_{\text{I}_R}$, baseline costs $C_{\text{B}_R}$,
triage costs $C_{\text{T}_R}$, and incident response costs $C_{\text{IR}_R}$.
As shown in Table \ref{tab:resource_costs} these resource costs can also be sub-divided by resource type.
This specificity helps in estimating costs and in relating costs to IDS performance and attack status.

The sub-categories of resources considered include the following:
\begin{itemize}
 \item \textbf{Licensing} - In most cases this will either be free, fixed cost, or a subscription based cost covering some time period.  However, this also could potentially involve a cost per host, cost per data volume, or some other system.  This will be a significant cost in many cases.

\item \textbf{Storage} - This is one of the easier costs to estimate; this increases approximately linearly with data volume.  This will generally be a function of the number of alerts generated, or a function of time if more routine information is being logged, such as logging all DNS traffic.

\item \textbf{CPU} - The computational costs of analysis, after data is collected, will (hopefully) scale approximately linearly with the data volume.  This cost can vary based on algorithm, indexing approach, and many other factors.  This is a function of time, and does not generally depend on the number of alerts, unless considering some process that specifically ingests alerts, e.g., security information and event management (SIEM) systems. There are additional costs of instrumentation and collecting data, for example capturing full system call records will impose some non-trivial cost on the host.  Most end-users are not CPU bound under normal workloads, so this cost is minimal as long as it's under some threshold.
In a cloud environment, this may be included in their billing model, or if self-hosting this will reduce the ability to oversubscribe resources, so in either of these cases the costs will be more direct.

\item \textbf{Memory} - There is some memory cost required for analysis and indexing.  This is generally a function of time, or in some cases a function of the number of alerts. 
There is also some memory cost for collection on the host.  Like CPU costs on the host, most physical machines are over-provisioned, so costs are minimal if under some threshold.  In a cloud environment, this will typically be a linear cost per time.

\item \textbf{Disk IO} - These costs are generally a function of time and/or a function of the number of alerts.  
This cost is not a major concern until it passes some threshold where it impacts performance on either a server or the user's environment.

\item \textbf{Bandwidth} - Like Disk IO, these costs are a function of time and of the number of alerts.
This is also not a major concern until it passes some threshold that causes performance degradation.

\item \textbf{Datacenter Space} - While in practice this is a large up-front capital cost, it would typically make sense to consider any appliances as `leasing' space from the datacenter.
Optionally, the rate set may account for how much of the datacenter's capacity is currently used,
so that space in an underutilized datacenter is considered a lower cost.
In commercial cloud environments, this is not a directly visible cost, but is included in other hosting costs.

\item \textbf{Power and Cooling} - These costs are similar to the costs of datacenter space discussed previously, 
except that representing the costs as a function of time is more direct.
In most cases this is not a major concern, but it could be in some cases, and is included for completeness.
\end{itemize}

\begin{table*}[t]
\centering
\begin{threeparttable}
   \vspace{-.1cm}
   \centering 
    \caption{Resource Costs}

    {\small
  \begin{tabular}{lcccccccc}
     \toprule
   Notation (Cost)  & Licensing	& Storage	& CPU	& Memory & Disk IO & Bandwidth & Space & Power \\ 
    \midrule
    $C_{\text{I}_R}$  (Initial) & $c_1$ & $0$ & $0$ & $0$	& $0$ & $0$ & $0$ & $0$ \\ 
     $C_{\text{B}_R}$ (Baseline) & $x_1 t$	& $x_2 t$ & $x_3 t$ & $x_4 t$ & $x_5 t$ & $x_6 t$ & $x_7 t$ & $x_8 t$ \\ 
     $C_{\text{T}_R}$ (Triage) & $0$ & $y_1n$ & $0$ & $0$ & $y_2n$ & $y_3n$	& $0$ & $0$ \\
    $C_{\text{IR}_R}$ (Incident Response) & $0$ & $0$ & $0$& $0$ & $0$ & $0$ & $0$ & $0$ \\ 
    \bottomrule
    \end{tabular}%
    } 
    \label{tab:resource_costs}%
\begin{tablenotes}[para]
\small
Table of resource costs for each component of $C_{\text{R}}$, as described in Section \ref{sec:resource_costs}.
\end{tablenotes}
\end{threeparttable}
\end{table*}

Initial costs $C_{\text{I}_R}$ would primarily consist of licensing fees and hardware purchases, as needed.
Hardware purchases and related capital costs, such as datacenter capacity,
can be either included in the initial costs or averaged over their expected lifespan,
which would be captured in the baseline costs $C_{\text{B}_R}$.
Either is acceptable, as long as they are not over- or under-counted.

Baseline costs $C_{\text{B}_R}$ represent the cost of normal operation,
when no alerts are being generated.
This may include licensing costs, if those are on a subscription basis.
This also would often include storage, CPU, memory, datacenter costs, and related costs,
in cases where hardware costs are amortized over time,
or in cases where cloud services are used, and these resources are billed based on usage.
This case is what is shown in Table \ref{tab:resource_costs}.

Alert triage costs $C_{\text{T}_R}$ represent costs of servicing and triaging alerts,
above the baseline costs of normal operation.
This is potentially a labor-intensive process,
but generally imposes little or no direct resource costs in terms of CPU, memory, etc.
The amount of storage and bandwidth needed for each alert is extremely small,
and is not significant until alert volumes become much higher than analysts could reasonably handle. 
There are some exceptions, such as large volumes of low-priority alerts,
or unusual licensing arrangements, so these costs are included for completeness.

Incident response costs $C_{\text{IR}_R}$ represent the costs of actually responding to a known attack.
Like the triage costs, this is labor-intensive,
but involves little or no direct resource costs outside of highly unusual circumstances.
This is included here simply for completeness.

%% file: 33-full-model.tex
\subsection{Full Model \& Parameter Analysis}
\label{sec:full-model}

Following this discussion in Section \ref{sec:defense-cost}, we can now replace $C_{\text{Defense}}$ with these terms, and represent $C_{\text{Total}}$ as the following:




\begin{equation}
\label{eqn:total-costs-expanded}
\begin{aligned}
C_{\text{Total}} = & \sum_{i} C_{\text{Breach}_i}
    + C_{\text{I}_L} + C_{\text{I}_R} 
    + C_{\text{B}_L} + C_{\text{B}_R} \\
    &\qquad + C_{\text{T}_L} + C_{\text{T}_R}
    + C_{\text{IR}_L} + C_{\text{IR}_R}
\end{aligned}
\end{equation}

As with all cost-benefit models, the primary downfall is estimating input parameters; e.g., populating $C_{\text{Breach}}(t)$ requires estimating the full impact of a future breach over time, an inherently imprecise endeavor.  
While we give some defaults and examples for many of the estimates in Sections \ref{sec:quantifying_costs} and \ref{sec:examples}, here we give a broad overview of sensitivity of the model to the parameters allowing users to target estimation efforts to those inputs that are most influential. 

Terms $C_{I_R}$ and $C_{I_L}$ are constants; hence, unless for some particular situation they are very large, they will not cause large effect when estimating costs over long time spans. 
Ongoing costs, $C_{B_L}$ and $C_{B_R}$ are linear, increasing functions of time. 
These will generally have  a greater effect than the constant one-time costs. 
In some cases these can be an outstanding contributor, but for most applications we expect them to be less influential than attack, triage, response costs. 

Triage costs, $C_{T_L}$ and $C_{T_R}$, are linear, increasing functions of the number of alerts, and incident response costs,  $C_{IR_L}$ and $C_{IR_R}$ are linear, increasing functions of the number of incidents and their cost. 
These are potentially very influential on the final costs. 
We note importantly that hidden variables are the false positive and true positive rates/quantities. 
The final costs of a security measure can vary widely with quantities of alerts and the accuracy of detectors, so these terms are very influential. 
This is supported by our examples where costs incurred by the quantity of false positives drastically vary overall costs. 

Finally, $C_{IR_L}$ and breach costs (attack models) are potentially non-linear in time. Consequently, they are the most influential parameters, along with hidden parameters ``how often do we expect to be attacked?'' and ``what type of attacks do we expect?'' 
As a quick example, the Ponemon's 2018 Report \cite{ponemon-cost-2018} gives statistics for breach costs, but also separate figures ``mega breach costs'' with the difference being two orders of magnitude in cost.  
Changing an attack or response model based on these two different estimates could potentially change total costs on the order of \$100M!

In summary, for most applications,  estimates of attack, incident response and triage costs will be most influential parameters. 
Importantly, estimating these requires latent variables such as true/false positive rates, which are in turn very influential.

%% file: 34-quantifying-costs.tex
\subsection{Quantifying Costs}
\label{sec:quantifying_costs}


When cost data or information on the effects of actual attacks are  available, the cost model's parameters can be computed relatively precisely.
When this data is not available, such as when evaluating a new product or scoring a competition, general estimates are available using prevailing wage information, cloud hosting rates, and similar sources.
To aid in application of the model, this  section  provides examples and reference values for the cost models introduced earlier in the section.


\subsubsection{Breach costs}
\label{sec:quant-breach-cost}
The costs of breaches ($\sum C_{\text{Breach}_i}$) can be estimated based on historical data, data aggregated from other organizations, and an estimate of the value of the data being protected.
For example, if a single host is infected with ransomware,
it may simply need to be re-imaged,
and the cost of this may be simple to estimate from labor costs to reimage a host (assuming that no data was exfiltrated as part of the attack).
If a more advanced adversary can infiltrate the network,
and they persist for long enough to find and exfiltrate valuable data, the cost of the breach rises dramatically after the adversary begins to steal data.
Modeling costs of such an attack will require estimates of the worth of the data in the organization and/or can rely on historical reports of similar breaches.

\begin{figure}[ht]
\centering
\includegraphics[width=.7\linewidth]{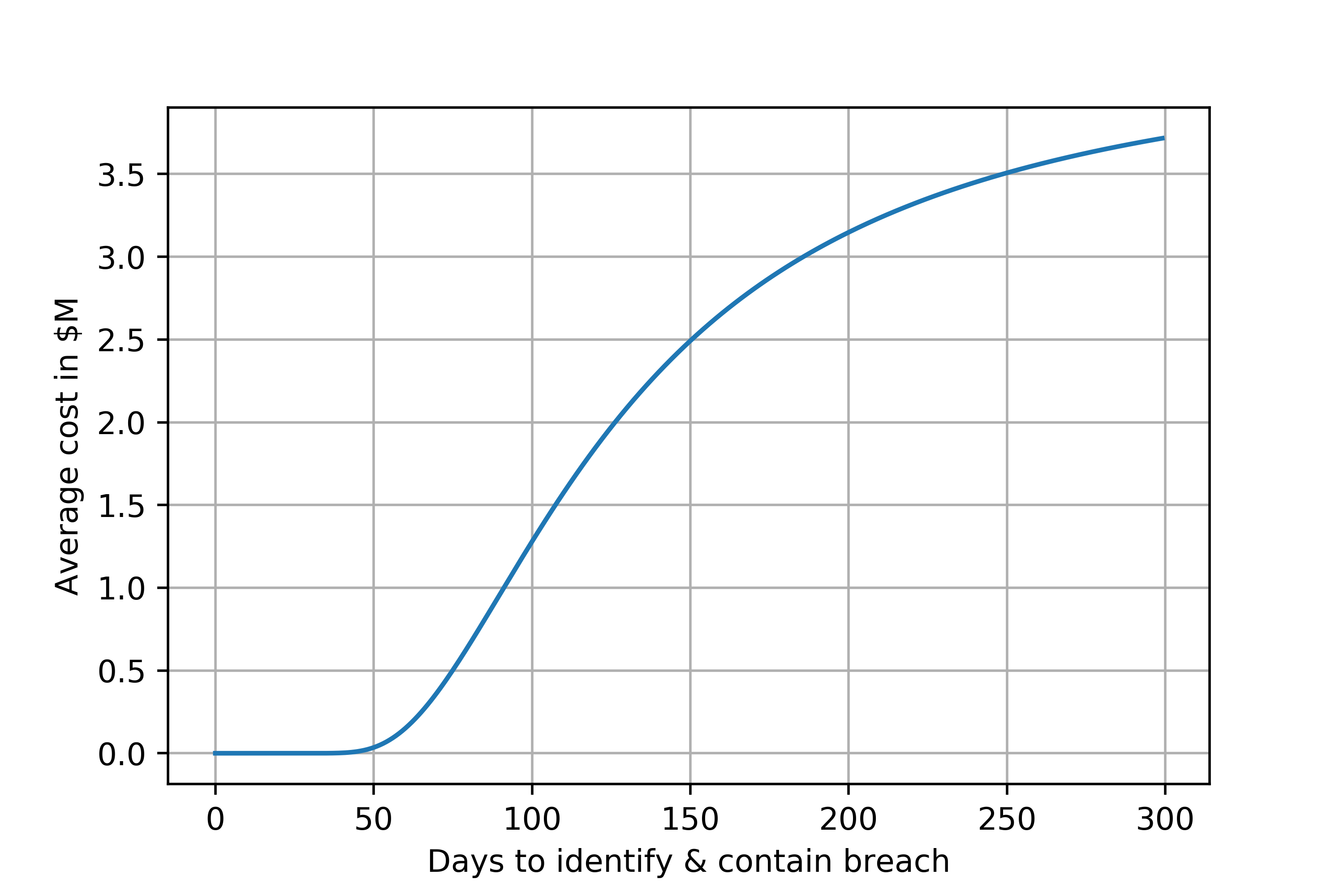}
\caption{Plot of $f(t) = 4.25 \exp(-12007.3/t^2)$, the S-curve estimate of attack cost given assumptions derived from Ponemon's 2018 data \cite{ponemon-cost-2018}. }
\label{fig:ponemons-cost}
\end{figure}

Using reports on breach costs from 2018, we provide an example of how to estimate an S-curve model (Section \ref{sec:defining_attacks_and_breaches}) of costs induced by an attack.
Ponemon's Institute provides a yearly report giving statistics on data breach costs and related statistics \cite{ponemon-cost, ponemon-resilience, ponemon-cost-2018}.
From the 2018 report we find ``The mean time to identify [a breach] was 197 days,'' and containing a breach in less than 30 days resulted in an average \$3.09M cost, while containment taking greater than 30 days cost \$4.25M.
We use these facts to fit  $f(t) = b \exp(-\alpha/t^2)$, the cost in \$M of a breach given discovery and containment occurred at $t$ days.
As no statistics are given about the distribution of time to discovery, we use the given average, 197 days as a default detection time in our calculations.
From the statement that containment taking greater than 30 days cost \$4.25M we obtain
$$
\lim_{t\to\infty} \frac{\int_{197+30}^{197+30+t} f(x) dx}{t}
 = 4.25.$$
For large $t$, $f(t)\approx b$, hence the limit on the left approaches $b$, giving $b = $\$4.25M.
Next, from the second piece of data we have
$$
 \frac{\int_{197}^{197+30} f(x) dx}{30}
 = 3.09.
 $$
 By numerically solving we can obtain $\alpha \approx 12007.3$. \\
Altogether our fitted S-curve breach cost model is \\
$f(t) = 4.25 \exp(-12007.3/t^2)$.  See Figure \ref{fig:ponemons-cost}.

Further examples of attack cost estimates are given in Section \ref{sec:virtue}.

\subsubsection{Resource costs}
\label{sec:quant-resource-costs}

\begin{table*}[t]
\centering
\begin{threeparttable}
   \vspace{-.1cm}
   \centering 
    \caption{Estimates of Resource Costs}

    {\small
    \centering
    \begin{tabular}{@{}p{\textwidth}@{}}
    \centering
    \begin{tabular}{lccc}
    \toprule
   \,\,\,\,Notation (Cost) 	& Storage	& CPU \& Memory & Bandwidth \\
    \midrule
    \,\,\,\,
    $C_{\text{I}_R}$  (Initial) & $0$ & $0$ & $0$  \\
    \,\,\,\,
    $C_{\text{B}_R}$ (Baseline) & \$0.10 per GB per month &
     \,\,\,\, \$20 per instance per month \,\,\,\, & $0$ \\
    \,\,\,\,
    $C_{\text{T}_R}$ (Triage) & \$0.10 per GB per month & $0$ &
     \,\,\,\, \$0.09 per GB out \,\,\,\, \\
    \,\,\,\,
    $C_{\text{IR}_R}$ (Incident Response)
    \,\,\,\, & $0$ & $0$ & $0$ \\
    \bottomrule
    \end{tabular}
    \end{tabular}%
    } 
    \label{tab:resource_costs_estimates}%
\begin{tablenotes}[para]
\small
Table of estimated resource costs, assuming and only taking into account cloud hosting fees.  This is calculated assuming long-term use of reserved t3.large AWS instances with EBS SSD at current (2018) prices.
For real-world scenarios $C_{B_R}$ Bandwidth would include prices for internet service, etc, and subscriptions depending on data volume (e.g. Splunk SIEM tool) fees would need to be added.
\end{tablenotes}
\end{threeparttable}
\end{table*}

Many of the resource costs described in Section \ref{sec:resource_costs} involve datacenter and hosting costs.
If these are unknown for a particular organization, calculating pricing based on cloud hosting provides a real-world default for these costs.
These costs are readily available from cloud hosting providers such as Amazon Web Services (AWS)\footnote{https://aws.amazon.com/pricing/}.

As shown in Table \ref{tab:resource_costs_estimates}, these costs primarily depend on the volume of data generated, and the amount of computational and memory resources needed for processing this volume of data.
Table \ref{tab:resource_costs_estimates} only includes AWS bandwidth costs, which does not charge for uploading data to their cloud services.
This does not include any costs from the ISP or other bandwidth costs.
Of course in a real-world scenario, a price for uploading data would be incurred, although not by AWS, but through bills for internet service, power, etc.
This also assumes no software licensing costs---some licenses, such as Splunk (\url{https://www.splunk.com}) a popular SIEM system, can significantly increase in cost based on data volume.

As noted in an earlier study of SOCs \cite{DBLP:journals/corr/abs-1812-02867}:
\begin{quote}
Reported size of host data varied widely ... On the low end an approximation of 300MB/day were given. One respondent works across many organizations and reported 100GB to 10TB per day, with the latter the largest estimate given during our surveys ...
Overall, Splunk subscription costs were cited directly by some as the constraint for data collection after mentioning they would benefit from more data collection. Perhaps this is unsurprising given estimates from the numbers above---the sheer quantity of host data collected and available to security operator centers is between 1GB-1TB/day, stored for 3 months $\approx$ 100days = 100GB-100TB.
\end{quote}

Combining the figures in the quote above with the costs from Table \ref{tab:resource_costs_estimates} results in baseline storage costs ranging from \$10 to \$10,000 per month; this shows that even though storage and bandwidth costs are very low per unit, they can be substantial across a large organization depending on what sources are collected.


\subsubsection{Estimates of Labor costs}
\label{sec:quant-labor-costs}
Current estimates place the average analyst salary at around \$75k to \$80k per year\footnote{See \url{www.glassdoor.com} salaries for ``Cyber Security Analyst'' and ``Information Security Analyst'' titles.} (about \$35 to \$38 per hour).
This figure does not include benefits, which generally make up 30 or 40 percent of total compensation\footnote{See \url{www.bls.gov/news.release/pdf/ecec.pdf} Table A.}, does not include any bonuses or other non-salary compensation, and does not include any overhead costs.  In the absence of any more detailed information, an estimate of \$70 per hour may currently be a reasonable starting point when including benefits and allowing some padding for other overhead.

Our interaction with SOC operators indicates that tens of thousands of alerts per month are automatically handled (e.g., an AV firing and quarantining a file), but a much smaller minority require manual investigation, usually tracked through a ticketing system.
A typical ticketed alert requires several minutes to triage by tier 1 analysts, and if escalated can require hours (or potentially days) to fully investigate and remediate according to some published sources \cite{zimmerman2014ten, sundaramurthy2016turning}.
Our interaction with SOC operators indicated that tier 1 analysts spend about 10 minutes per ticketed alert, that tier 2 analysts use up to 2 hours, and tier 3 analysts time is potentially unbounded.
If we assume that 50\% of alerts can be triaged and resolved by tier 1 analysts\footnote{In reality, this figure may be significantly lower; we consulted a few SOC operators who reported about 90\% of ticketed alerts advance to tier 2 analysts.  This is highly dependent on the organization, the source of the alerts, and the false positive rate.}, in an average of 10 minutes, and that additional investigation by higher tiers takes an average of 2 hours, that means an average alert would cost approximately 10 minutes at \$70/hr (\$11.67) $+$ 50\% $\times$ 2 hours $\times$ \$70/hr (\$70), or about \$80 in total.

After triage, the incident response begins, i.e.,
handling cleanup, mitigation, and related tasks after an attack.
Section \ref{sec:labor_costs} proposes an S curve of increasing cost over time, and with some information on costs of incident clean up one could fit an S curve similar to the example in Section \ref{sec:quant-breach-cost}.
For a simpler model, if we assume an average of 6 hours for incident response at \$70/hr, we obtain a cost of \$420 or about \$400.


\begin{table}[t!]
\centering
\begin{threeparttable}
   \vspace{-.1cm}
   \centering 
    \caption{Estimates of Labor Costs}
    {
    \begin{tabular}{l c }
    \toprule
    Notation (Cost) & Analysts \\
    \midrule
    $C_{\text{I}_L}$ (Initial) & $0$  \\
    $C_{\text{B}_L}$ (Baseline)  & $0$  \\
    $C_{\text{T}_L}$ (Triage) & \$80 per alert  \\
    $C_{\text{IR}_L}$ (Incident Response) & \$400 per incident \\
    \bottomrule
    \end{tabular}%
    } 
    \label{tab:labor_costs_estimates}%

\begin{tablenotes}[para]
\small
  Table of estimated labor costs for analysts.  This assumes little or no impact on end users, which may not be true in all organizations.
\end{tablenotes}
\end{threeparttable}
\end{table}

%% file: 40-examples.tex
\section{Examples}
\label{sec:examples}
Here we provide specific examples of the using the framework. 
The first example explains estimates and configuration of the framework for an upcoming IARPA grand challenge, a cyber competition and the target application driving this research. 
Secondly, we evaluate a detection algorithm proposed in our previous work as though it was to be deployed. The second application gives examples of how the evaluation framework is useful from the point of view of the researcher in developing novel tools/algorithms, from the SOC in considering purchase of a new tool, and from a vendor deciding the worth of their product.

%% file: 41-virtue.tex
\subsection{VirtUE Contegrity Breach Detection Challenge}
\label{sec:virtue}
A target application of this framework is evaluating detection capabilities for a competition as part of the  IARPA VirtUE (Virtuous User Environment) research and development (R\&D) program\footnote{\url{https://www.iarpa.gov/index.php/research-programs/virtue/virtue-baa}}. 
The VirtUE R\&D program is developing a computing environment where each of a user's daily computing roles occupies its own isolated virtual environment (a VirtUE) without significant impact on the functionality to a user, 
e.g., separate Docker \cite{merkel2014docker} containers could be launched for a user's email, internet browsing, and Sharepoint administration roles, while the user sees and interacts with a single unified desktop presenting all these roles. Building isolated virtual environments specifically for constrained, well-defined user roles creates enhanced opportunities to sense and protect those environments. VirtUE hopes to contrast this with the traditional user interface model where all user roles are merged indistinguishably into one single shared-memory environment.

In the VirtUE Contegrity Challenge, competitors  are tasked with accurately identifying attacks on confidentiality and integrity (contegrity) as efficiently as possible.  
Specifically, the competitors will employ their detection analytics to analyze the security logs of six different role-specific Virtues. Competitors will be tasked with minimizing the total amount of log data that their analytics process while accurately detecting the presence of contegrity attacks on a Virtue. Each Virtue will experience  zero to two attacks over a time period of an hour for a total of 12 possible attacks. 
The attacks fall into 16 categories (e.g., ``Capturing or transporting encryption keys'',  ``Corrupting output of a computation''), and performers must identify the class of the attack with each alert.  
Alerts  with the wrong classification are considered a false positive.

\subsubsection{Competition Scoring Model} 
The goal of this section is to produce a scoring procedure that accomplishes the following: 
\begin{itemize}[itemsep=-2pt]

\item rewards accuracy 
of the detector  
\item rewards timeliness of detection
\item penalizes for bandwidth, processing, memory, and storage use
\item is practical to compute for evaluating such a competition
\end{itemize}
In short, the scoring should take into account the accuracy, timeliness, and resource requirements of the detection capability. 
We leverage the model above to determine a ``cost of security'' score for each participant's detector and the detector incurring the lowest cost wins. 

\begin{figure}[ht]
\centering
\includegraphics[width=.7\linewidth]{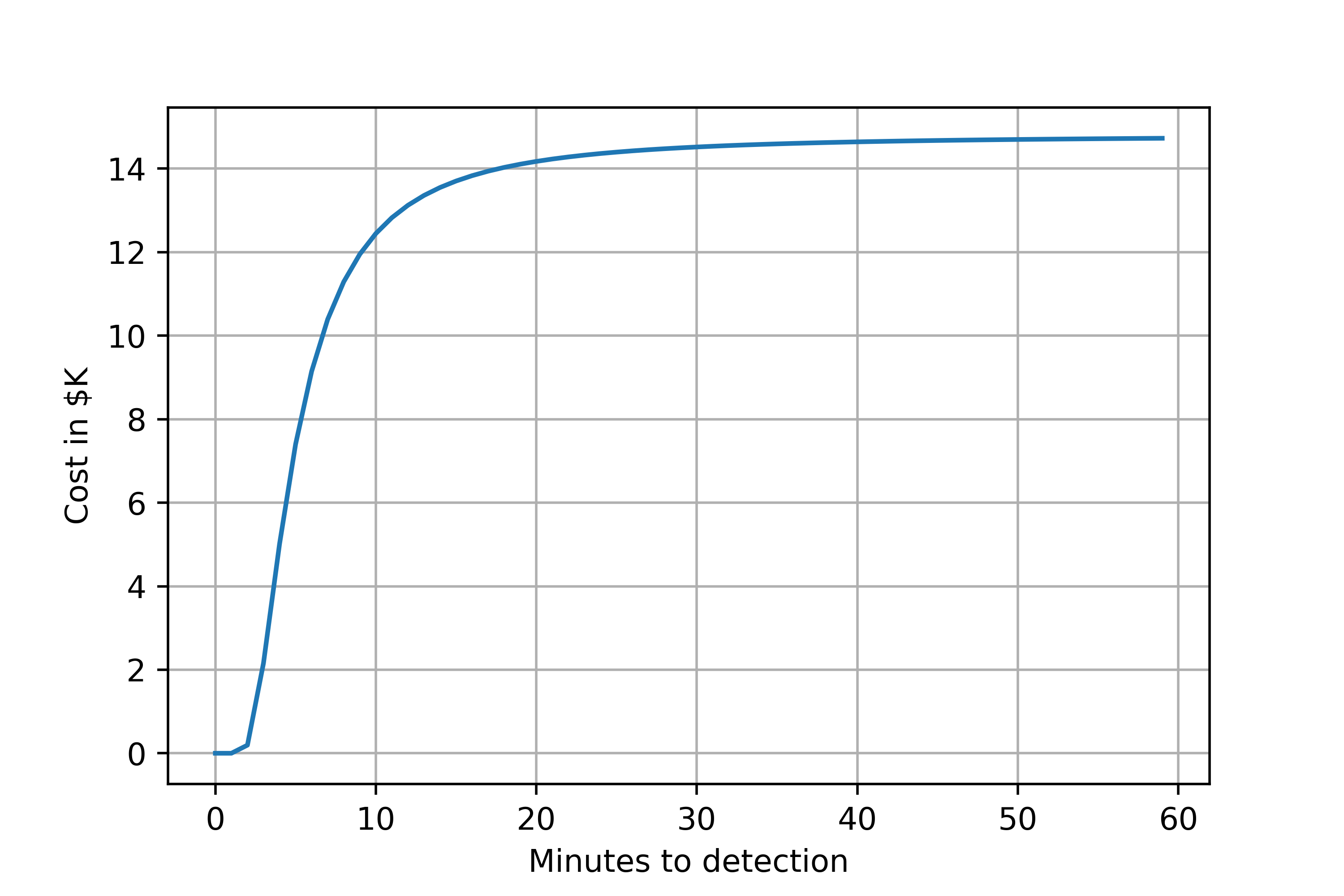}
\caption{Plot of S-curve $f(t) = 14.8\exp(- 25 \log(2)/t^2)$, the  estimate of attack cost for VirtUE challenge.}
\label{fig:virtue-cost}
\end{figure}

To model the attacks, we assign a total value of the data each Virtue contains, and this provides the asymptote  ($b$) for the S-curve model as described in Section \ref{sec:breach_costs}.  
Consulting Ponemon's 2018 report \cite{ponemon-cost-2018}, the average cost per client record affected in a breach was \$148 in 2018. 
(We note that Ponemon's report focused on stolen customer data, which may not be an accurate estimate for enterprise data.)  
Assuming 100 files per Virtue furnishes $b =$ \$14,800. 
To accommodate the 1-hour competition duration, we choose the time parameter, $\alpha$, so that 50\% of the maximum possible cost, $b$, is obtained in 5 minutes. 
Thus, $\alpha = 25 \log(2)$.  
Note that while we model integrity attacks with equal cost as confidentiality attacks, in an alternate scenario one may adjust the model to give greater cost to attacks against integrity.  This is because an adversary with write access to some data typically also has read access to that data, so attacks against integrity should be equal or greater in cost.
Altogether, our competition's model for the cost of an integrity attack (in thousands of dollars) $t$ minutes after initiation is
\begin{equation}
\label{eqn:virtue-attack}
    f(t) = 14.8e^{- 25 \log(2)/t^2}.  
\end{equation} 
See Figure \ref{fig:virtue-cost}. 
Finally, for each attack administered in the competition, we charge the participant $f(t_d)$ thousand dollars, where $t_d$ is the attack duration lasting from the start time of the attack to either the time of correct detection or the end of the 1-hour competition. 
Note that since $f$ is increasing, this rewards early detection over later. 

For labor and resource costs, we follow Section \ref{sec:quantifying_costs}, namely Tables \ref{tab:resource_costs_estimates} and  \ref{tab:labor_costs_estimates}, with some tweaks. 
As the goal is to evaluate the efficacy of the detection algorithms, we can ignore licensing and configuration costs for the detection software ($C_{I_R} =$ \$0, $C_{I_L}$ = \$0), which is equivalent to assuming each competitor's software incurs the same licensing and configuration costs,  as well as baseline labor costs ($C_{B_L} =$ \$0) and resources needed for incident response ($C_{IR_R} =$ \$0).   
Competing detection capabilities will be furnished a uniform CPU and memory platform, hence CPU and memory costs can be ignored. 

As explained in Section \ref{sec:quant-resource-costs} every alert will cost \$80 to triage, regardless of whether the alert is a true positive or false positive.  
If the alert is a true positive, the attack is considered detected and remediated, and ceases to accrue cost; however, a fixed \$400 fee is incurred to represent the cost of this remediation. 

Ongoing resource costs for use ($C_{B_R}$) and triage costs ($C_{T_R}$) will be traceable during the competition. 
To estimate these costs, we monitor the volume of data sent in or out of the detector's analytic environment  and  charge a single per-volume rate of \$150/MB to account for any bandwidth, storage, or per-volume subscription fees (e.g., SIEMs). 
Just as the expected number and time-duration of attacks (and therefore alert costs) are condensed to accommodate the 1-hour duration, we inflated the estimated data costs to make it comparable to the attack and alert costs expected.  

Here we itemize this estimate. 
SOCs generally store logs and alerts for at least a year (e.g., see  \cite{DBLP:journals/corr/abs-1812-02867}). 
This requires movement of data to a datastore,  storage fees for a year,  and SIEM license fees. 
We estimate bandwidth costs at \$1/GB based on the low estimate of \$0.09 cloud bandwidth fees (Table \ref{tab:resource_costs_estimates}) and high estimates from mobile networks (e.g., needed by shipping vessels and deployed military units) that can cost \$10-\$15/GB. 
For 1 year of storage we consult the cloud costs in Table \ref{tab:resource_costs_estimates} and obtain \$0.10/GB/month $\times $12 months  = \$1.20/GB for 1 year of storage. 
For storage and management software price, we reference Splunk, costing \$150 / GB / month.\footnote{Price from  \url{https://www.splunk.com/en_us/software/pricing.html} as of 09/25/2018.} 
Our estimate is \$150 / GB / month $\times$ (1/730.5) month / hour = \$0.20 / GB / hour for the portion of the SIEM fee incurred in this 1-hour competition. 
Altogether, a reasonable estimate for data moved in or out of the detector's environment is \$3.40/GB, comprised of: \$1/GB for the observed data movement, another \$1/GB assuming a copy of it is sent to long term storage, \$1.20/GB for storage fees, and \$0.20/GB for SIEM fees. 
Finally, we need to scale this price to be comparable to the condensed attack costs in the hour. 
The attack volume is on the order of the number of attacks expected of a single host in perhaps a calendar year, yet the competition involves only about a fifth of the Virtues needed for  a single virtual host.  
Consequently, to make the data costs comparable to the attack  and alert triage/remediation costs, we multiply by 5 $\times$ 24 hours/day $\times$ 365 days/year, giving \$148.92/MB. For simplicity we use \$150/MB in the competition.

Altogether, the scoring evaluation is as follows: 
\begin{itemize} 
\item When the competition starts, the system is in a true negative state, and only the bandwidth/storage used by the detector will be accruing costs. Total volume of data used by the detector's virtual environment will incur a fee of \$150/MB. 
\item Every time an alert is given an \$80 fee will be charged for triage.  
\item For every true positive alert, an additional \$400 fee will be charged to represent the cost of remediating the attack. 
\item Once an attack is detected, $\Delta t_d$, the time from the start of the attack until detection, is determined. The attack is considered ended and $f(\Delta t_d)$  thousand dollars is charged.   
\item Finally, at the end of the competition, for any ongoing (undetected) attacks their duration (from start of the attack until the end of the competition) $\Delta t_d$ is determined. 
For each, $f(\Delta t_d)$  thousand dollars is charged.   
\end{itemize}

\subsubsection{Testing the Virtue IDS Evaluation Framework: Simulations \& Baselines} 
Importantly, we seek confirmation that the scoring procedure does indeed reward a balance of accuracy, timeliness, and preservation of resources.  
To investigate, we simulate some attack scenarios and defense schemes and present the results. 
We create four separate scenarios with number of attacks, $n_a = $0, 4, 8, and 12, and the $n_a$ attacks occurring at randomly sampled times (rounded to nearest two minutes) in the hour with replacement. 


Our detection models are as follows: 
\begin{itemize}
    \item {\it Null detector - } This detector simulates having no security measures.  It uses no data, throws no alerts and has maximum time to detection  (60m - attack start time). 
    \item {\it Periodic hunting (10m) - } This detector and the one below simulates a full system check at preset intervals without continually monitoring any data. We assume it will detect any ongoing attacks in each scan  but will also incur many false positives by issuing all 16 alerts at 10m, 20m, ..., 50m. Data cost is \$0 as it does not monitor hosts. 
    \item {\it Periodic hunting (30m) - }Same as above but with only two scans at 15m and 45m. 
    \item {\it Low-data, low-speed detector - } This and the detectors below simulate a real-time monitoring IDS. 
    It uses 5 MB  of data (initial overhead) plus 0.1 MB per attack; $Data(n_a) = 5 + 0.1n_a$ MB. 
    It throws 5 alerts plus 5 per attack; $Alerts(n_a) = 5 + 5n_a$. (Hence we assume that with no attacks we obtain 5 false positives, and we assume that for each attack it sends an additional 4 false alerts then the correct fifth alert.)  
    It detects every attack at 3 minutes. 
    \item {\it Low-data medium-speed detector - } Same as the above detector, but we assume it detects every attack at 1.5m.  
    \item {\it High-data medium-speed detector - }
    It uses 10 MB of data plus 1 MB per attack; $Data(n_a) = 10 + n_a$ MB. 
    It throws  4 alerts plus  4 per attack; $Alerts(n_a) = 4 + 4n_a$. 
    It detects every attack at 1.5m. 
    \item {\it High-data high-speed detector - }
    Same as above but it detects every attack at 15s. 
\end{itemize}

\begin{figure}[ht]
\centering
\includegraphics[width=.9\linewidth]{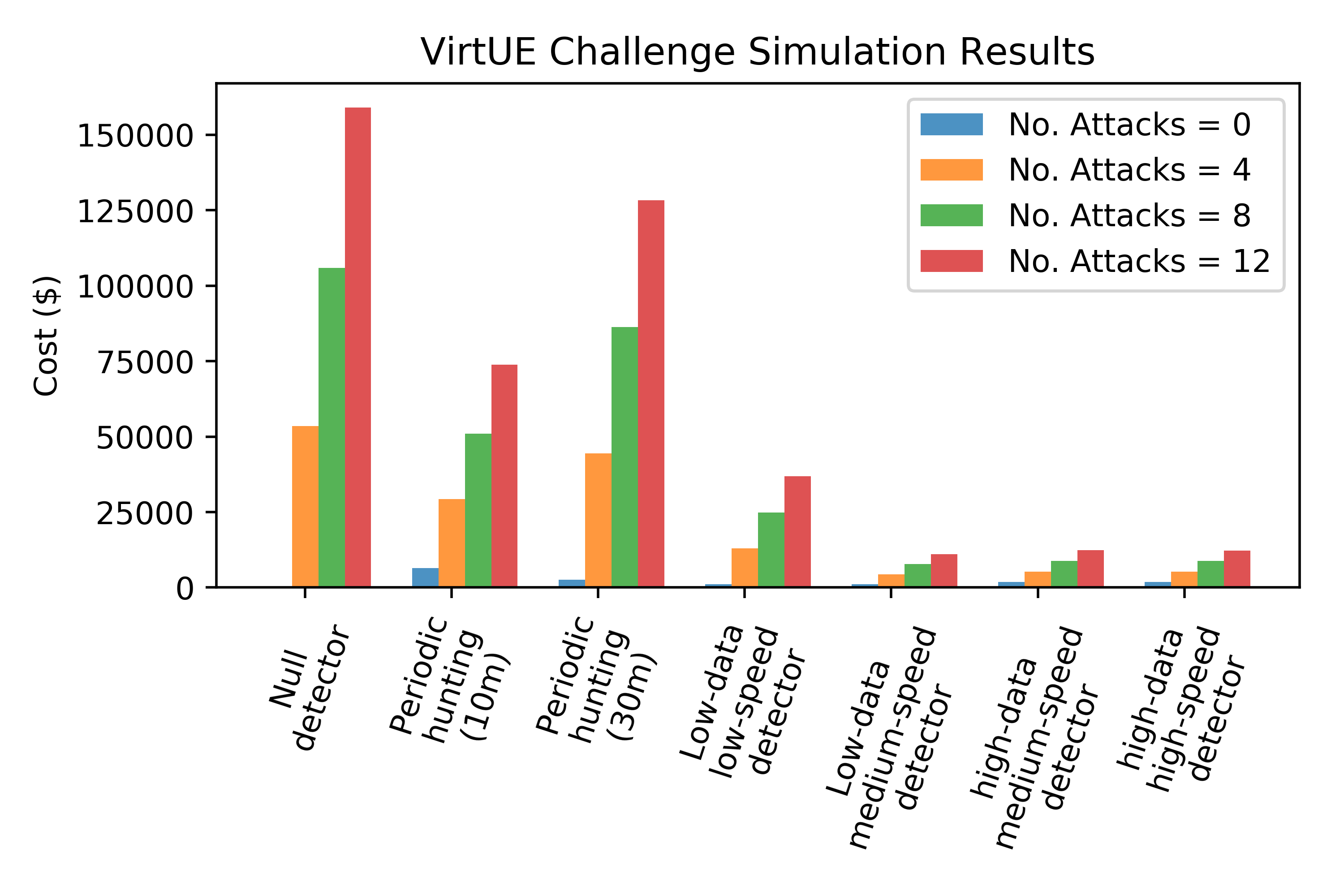}\\
\includegraphics[width=.9\linewidth]{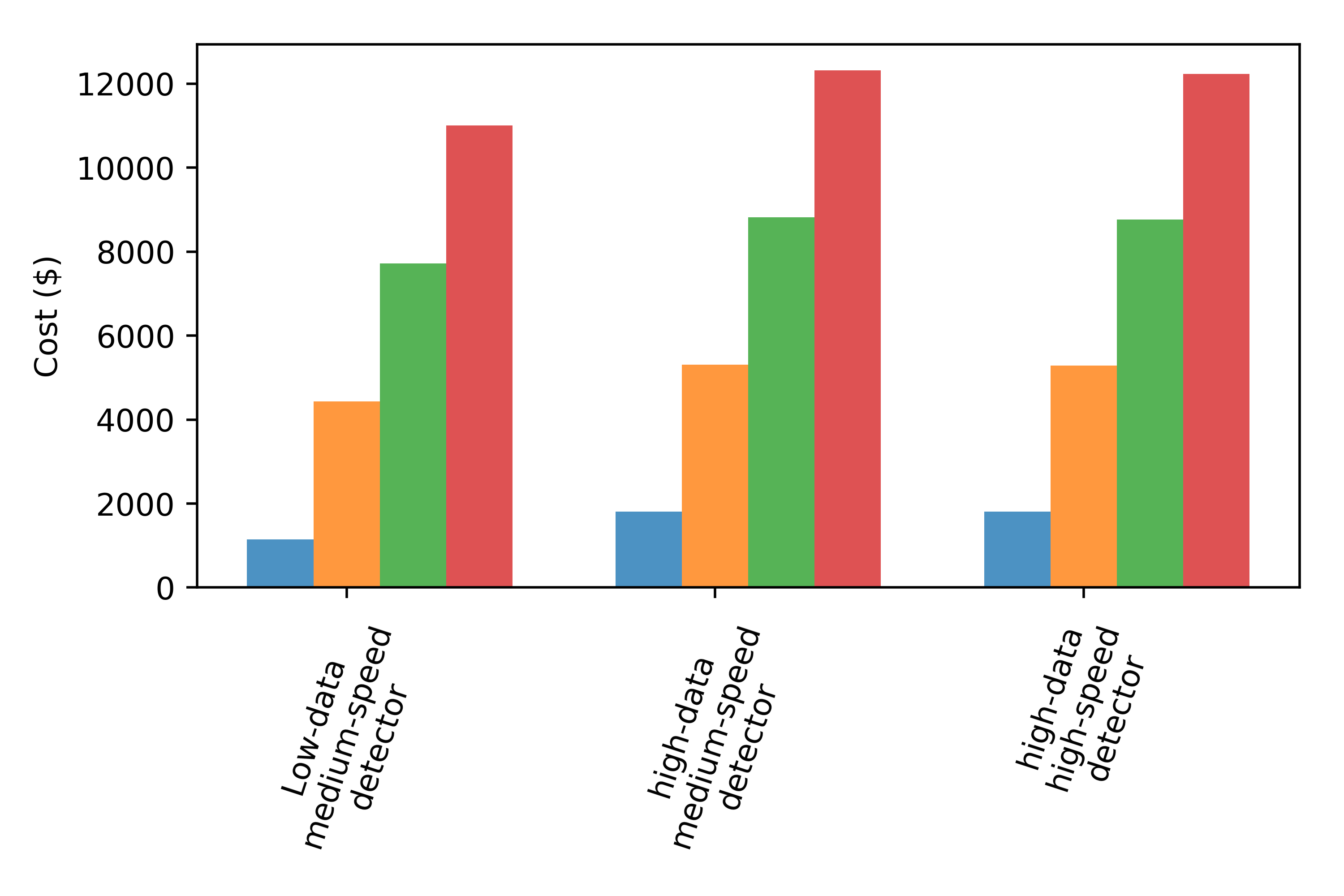}
\caption{For each scenario (No. Attacks = 0, 4, 8, 12), the cost incurred by each simulated detector was computed over 1000 runs and the average cost reported in the bar charts. 
Top bar chart presents the full results, while the lower bar chart is simply a zoom-in on the last three detectors. 
}
\label{fig:simulation}
\end{figure}

Results are displayed in Figure \ref{fig:simulation} giving the average cost incurred by each simulated detector over 1000 runs in each attacks scenario (0, 4, 8, 12 attacks). 
As expected, the Null Detector (no security measures) incurs no cost if there is no attack, but averages wildly high attack costs in the attack scenarios. 
Periodic hunting for threats incurs a large cost during the investigations, but strictly limits the attack costs. Performing the scans for attacks every 10m vastly outperforms periodic hunting on 30m intervals when attacks are present. 
We see a dramatic drop in costs across all scenarios with attacks when any real-time monitoring is assumed (last four detectors) even when only approximately 1 true alert in 5 was assumed. 
See the bottom bar chart for a zoom-in on the three best simulated detectors. 
First note that decreasing time to detection in both the low-data and high-data detectors also decreases costs, as desired. 
Further, as these four models increase linearly with the number of attacks, simply looking at the 12-attack scenario suffices. 
Next, note that the overall best performance is by the low-data, medium-speed detector, which takes 1m to identify an attack (\$11,010 for 12-attack scenario). 
We note that while the high-data, high-speed detector detects attacks six-times faster, its use of data increases its cost (\$12,260 for 12-attack scenario), but it is still slightly better than the high-data medium-speed detector (\$12,340 for 12-attack scenario) as expected. 
These nearly identical costs for the two high-speed detectors imply that detection within the first minute or so of an attack effectively prevents the attack; hence, it is not worth the data costs to increase time to detection in this case. 

Overall, these are comforting results because they suggest that simple heuristics for sending alerts without actually monitoring activity will incur too large a penalty in terms of false positive costs and attack costs to be as effective as intelligent monitoring.
This indicates that the scoring system is not easily `gamed'.
Further, this shows that this cost model requires a balance of data use, accuracy, and timeliness to minimize costs. 

We hope these simulations provide useful baselines for competitors. 

\subsubsection{Comparing Results with Other Evaluation Methods}
We now consider these simulated IDS strategies through the lens of previous quantifiable evaluation methods.
For all scenarios we assume 720 non-attack events occur (one each 15-second interval)---note that this is needed for computing accuracy metrics, but not for our cost evaluation as the number of false positive is defined in each detection strategy and true negatives incur no costs.

Table \ref{tab:detection-metrics} displays both the classical detection metrics and then an adaptation of the time-dependent metrics proposed by Garcia et al. \cite{garcia2014empirical} (see Section \ref{sec:ctu} CTU Botnet Dataset). 
Results are micro-averages across 1000 simulations. 
Recall that the CTU time-dependent detection metrics are computed each time window and encode that earlier detection or non-detection of attack traffic (true positive and false negatives) is time-sensitive, but normal traffic (true negative and false positive) is not impacted by time.
This is accomplished by computing the confusion matrix at each time interval according to Section \ref{sec:ctu}. 
To compute microaverages, the 720 confusion matrices are summed for each simulation, then summed over the 1000 simulation, and then the accuracy metrics are computed.  
To adapt their framework to this experiment we set parameter $\alpha = 1$, and used 15-second time windows. 
Note that once an attack begins, false negative scores are gained until it is detected, after which point true positive scores are gained; furthermore, since false positives only occur on the time intervals in which false alerts are produced, this scoring inflates precision and deflates the false positive rate. 
Further, we note that the actual scores produced are brittle depending on the number of time intervals used.

Beginning with the standard accuracy metrics our simulation produces best results for the high-data high- and medium-speed detectors. 
This is unsurprising as it neglects both the host resources used and the time to detection; hence, it promotes these two and has no means of distinguishing them. 
The CTU time-dependent metrics take into account time to detection, but not host resources. 
Hence, we see it champions the high-data, high-speed detector. 

\begin{table*}
\centering
\begin{threeparttable}
   \vspace{-.1cm}
   \centering 
    \caption{Standard and CTU Time-Dependent Accuracy Metrics}
    {
    \begin{tabular}{@{}p{\textwidth}@{}}
    \centering
    \begin{tabular}{l| c c c | c c c }
    \toprule
     & \multicolumn{3}{c|}{\textbf{Standard Metrics}} & \multicolumn{3}{c}{\textbf{CTU Metrics}}\\
    \textbf{Detection Strategy} & \,\,\,\,\textbf{TPR}\,\,\,\, & \,\,\textbf{FPR}\,\, & \textbf{Precision} & \,\,\,\,\textbf{TPR}\,\,\,\, & \,\,\textbf{FPR}\,\, & \textbf{Precision}  \\ 
    \midrule
    Null detector & 0.0 & 0.0 & N/A & 0.076 & 0.0 & 1.0 \\ 
    Periodic hunting (10m) & 0.864 & 0.097 & 0.13 & 0.794 & 0.001 & 0.999\\ 
    Periodic hunting (30m) & 0.759 & 0.032 & 0.285 & 0.563 & 0.000 & 0.999 \\ 
    Low-data low-speed detector & 0.967 & 0.106 & 0.178 & 0.879 & 0.008 & 0.991\\ 
    Low-data medium-speed detector & 1.0 & 0.09 & 0.185 & 0.922 & 0.008 & 0.992\\
    High-data medium-speed detector \,\,\,\, \,\,\,\, \,\,\,\, & 1.0 & 0.072 & 0.231 & 0.922 & 0.006 & 0.993\\ 
    High-data high-speed detector & 1.0 & 0.072 & 0.231 & 0.987 & 0.006 & 0.994 \\
    \bottomrule
    \end{tabular}
    \end{tabular}%
    } 
    \label{tab:detection-metrics}%
\begin{tablenotes}[para]
\small
  Table depicting the standard accuracy metrics and the CTU time-dependent metrics (Sec \ref{sec:ctu}). Results are microaverages over 1000 simulations. Note that the standard metrics promote the most accurate detectors (high-data *) as both time to detection and host resources are not respected. Whereas, the CTU time-dependent metrics favor the high-speed detectors that exhibit more accuracy and speed but neglect host resources used. Recall our metric promoted the low-data medium-speed detector as a balance of accuracy, host resources, and timely detection. 
\end{tablenotes}
\end{threeparttable}
\end{table*}

As a final baseline, we revisit the $1/e$ rule of the Gordon-Loeb (GL) Model \ref{sec:gordon-loeb}, which states that the optimal cost of security should be bounded above by the estimated loss to attacks over $e$. 
For the 12-attack scenario, the Null detector (no security) simulated costs was \$159,093. Dividing by $e$ gives the GL upper bound for optimal security costs of \$58,527.  
We note that both the periodic detectors are above this bound, while all four monitoring simulations are under it. 



%% file: 42-graphprints.tex
\subsection{GraphPrints Evaluation Example} 
\label{sec:graphprints}
In this section we revisit our previous work \cite{harshaw2016graphprints} that introduced GraphPrints, a network-level anomaly detector using graph-analytics. 
Our goal is to provide an example of the evaluation framework as an alternative to the usual true-positive/false-positive analysis given in the original paper and commonly used for such research works. 
Additionally, the example illustrates how the cost-benefit analysis can benefit (1) the researchers evaluation of a new technology, 
(2) SOC operators from the perspective of considering adoption as if GraphPrints were a viable commercial off-the-shelf technology and
(3) from the point of view of a vendor deciding on the price of such a product. 

\subsubsection{GraphPrints Overview} 
The GraphPrints algorithm uses network flow data to build a directed graph from each time slice (e.g., 30s.) of flows. 
The graph's nodes represent IPs and directed, colored edges represent connections with port information.

{\it Graph-level detection:} For each graph the number of graphlets---small, node-induced sub-graphs---are counted. This gives a feature vector encoding the local topology of the communications in that time window. 
A streaming anomaly detection algorithm is performed on the sequence of graphlet vectors. 
Specifically, a multivariate gaussian distribution is fit to the history of observed vectors, and new graphlet vectors with a sufficiently low p-value (equiv. high mahalanobis distance)  are detected as anomalous. 
Finally, the newly scored vector is added to the history of observations, and the process repeats upon receipt of the next vector. 
This provides an anomaly detector for the whole IP space represented by the graph. 
We note that the original GraphPrints  paper \cite{harshaw2016graphprints} also describes a related, node-level detector (following Bridges et al. \cite{bridges2015multi}), but for the sake of clarity, we provide the evaluation for only the network-level technology. 


\subsubsection{GraphPrints Evaluation}
For testing in the original paper \cite{harshaw2016graphprints}, real network flow data was implanted with bittorrent traffic as a surrogate for an attack. 
As torrenting was against policy it indeed constituted anomalous traffic. 
Secondly, it was chosen as bittorrent traffic appears as an internal IP contacting many external, abnormal IPs and moving data, potentially similar to malware beaconing or data exfiltration. 

\begin{figure}[ht]
\centering
\includegraphics[width=.99\linewidth]{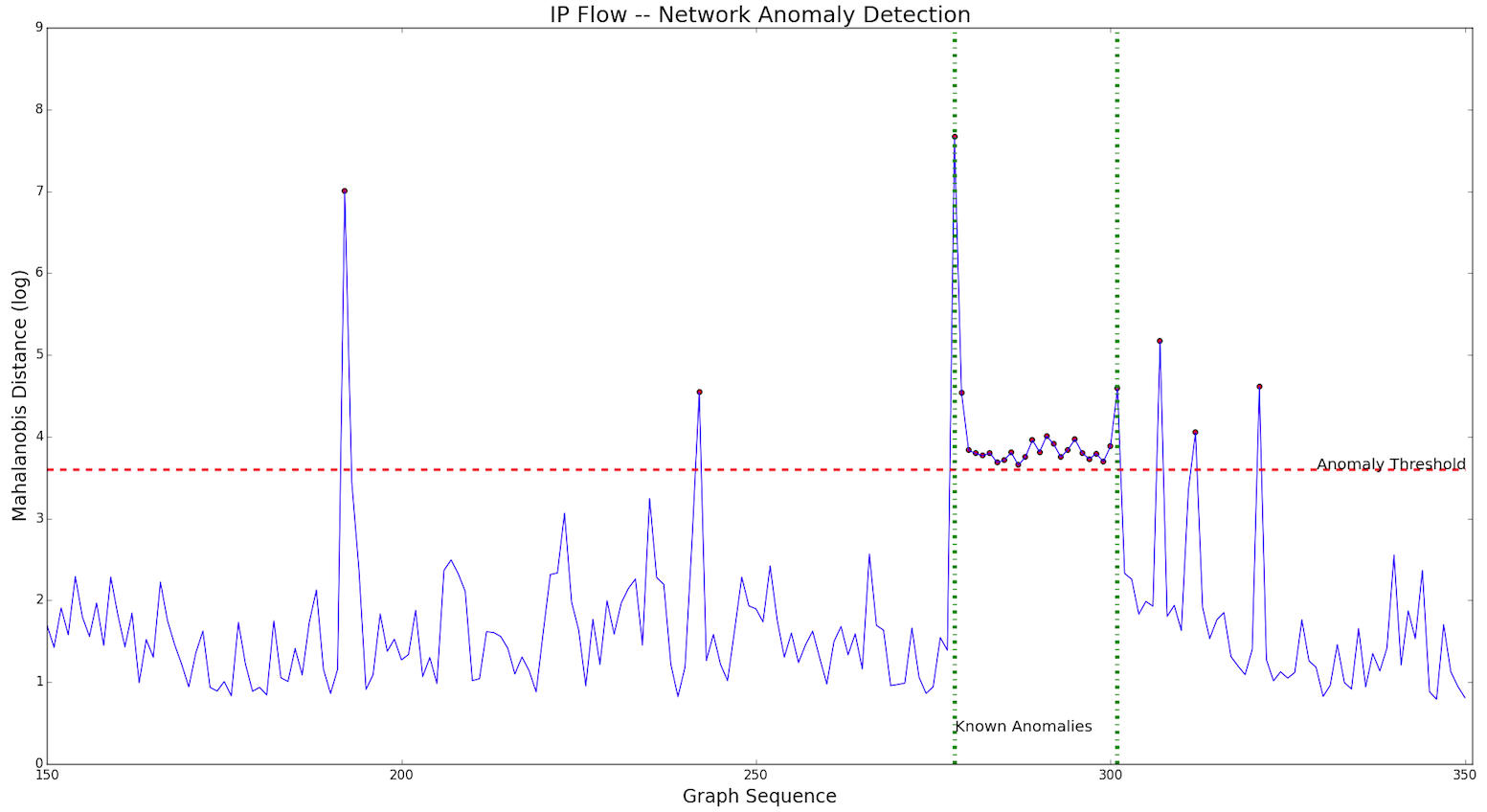}
\caption{\small{Original plot from GraphPrints paper  \cite{harshaw2016graphprints} depicting network-level anomaly scores. Suggested threshold is depicted by the horizontal, red, dashed line. Vertical green dashed lines indicate beginning/ending of simulated attack (positives). Spikes above the threshold, outside the attack period are false positives. }}
\label{fig:graphprints-network-results}
\end{figure}

The initial evaluation sought to show the existence of a window of thresholds for the  detector that gave ``good'' true/false positive balance. 
See Figure  \ref{fig:graphprints-network-results}. 
At the network level 
 with the depicted threshold the test exhibited perfect true positive rate and 2.84\% false positive rate. 
We manually investigated the false positives finding they were IP-scans originating from internal hosts assigned to the company's IT staff. 
Presumably this was legitimate activity causing false positives, e.g.,  a vulnerability or asset scanning appliance. 

We provide an instantiation of our evaluation framework as a more informative alternative to the true/false positive analysis of the detection capability. 
To estimate the initial resource cost, $C_{I_R},$ the cost of necessary hardware is tallied. 
Based on preliminary testing we conducted, to run the algorithm in real time a separate instance should be used to model roughly 2,500 IPs. 
That is, we expect a large network to be divided into subnets with separate GraphPrints instances per subnet,  
e.g., an operation w/ 10,000 IPs would require $n = 10,000/2,500 = 4$ GraphPrints servers. 
Since all costs except the initial subscription will scale by $n$ we  neglect this factor in the analysis and note that the final figures grow linearly with the network size. 
We contacted a few SOCs regarding server specifications for such a technology, and they pointed us to Thinkmate HPC\footnote{\url{www.thinkmate.com/systems/servers/rax}} and Cisco UCS C220 M4 rack server\footnote{\url{www.cisco.com/c/en/us/products/servers-unified-computing/ucs-c220-m4-rack-server/index.html}} costing approximately between \$2K to \$15K depending on configuration options. 
Additionally they mentioned adding 15\% for un-included hardware, e.g., racks, cords, etc. 
Most software used is opensource (e.g., Linux OS). 
Altogether, we estimate  $C_{I_R} =  \$7,500 \times (1.15) = \$8,625 $ per instance. 
Additionally, the initial labor costs to configure the servers we estimate at one day giving $C_{I_L}$ = \$70/hour $\time 8$ hours = \$560, following rates estimated in Section \ref{sec:quant-labor-costs}.  

For $C_{B_R}$ we assume the operation already collects and stores network flows, so adding this technology will add only alerts to the storage costs, and will add flows and alerts to the bandwidth cost, as alerts are sent to the SIEM and flows must be sent from the flow sensor to the GraphPrints server. 
For storage costs, if we assume 500KB/month/server, estimates in Table \ref{tab:resource_costs_estimates} give $500 \times 10^{-6}$ GB/month $\times $\$0.10/GB = \$$5 \times 10^{-5}$/month. 
Since this is a negligible amount of money, we ignore this term in the estimate. 
For bandwidth costs,  assuming \$0.10/GB, a low-end estimate for data transfer since it is internal, we estimate about 15GB of flows are produced per subnet per month, giving \$1.5/month. 
This is again, a negligible amount comparatively, so we ignore $C_{B_R}.$

We estimate that each  instance of GraphPrints will require weekly reconfiguration, e.g., threshold adjustment or a heuristic implemented to reduce false positives, and we allocate 1 person hour per week per instance. 
From estimates in Section~\ref{sec:quant-labor-costs}, $C_{B_L} = $\$70 / week $\times $ 4 weeks / month = \$280 / month. 

For triage and incident response costs, we reuse the estimates from the VirtUE challenge above;  namely, we assume a flat average of \$80 per false positive, and \$480 per true positive. 
Similarly, for breach costs we use the S curve fron the VirtUE challenge given in Equation ~\ref{eqn:virtue-attack}. 
Given the results above (Figure \ref{fig:graphprints-network-results}) we assume a perfect true positive rate with near immediate detection (assume, containment within 1 minute response time), and a 2.84\% false positive rate. 
For each GraphPrints instance, a scored event occurs every 30s time window giving 86,400 events $\times$ 2.84\% = 2,454 false positives per month, accruing 2,454 $\times$ \$80 $ = $ \$196,320/month. 
If we assume each instance incurs one attacks per month,
then a 1-minute response time gives response + attack costs of \$480 + $f(1)$ = \$480 / month (attack costs are negligible with fast detection).  

Altogether, the cost of adopting this technology, neglecting licensing or subscription fees is estimated as an initial one-time resource and configuration fee of \$8,625$+$560 = \$ 9,185 and a ongoing monthly cost of \$ (196,320 + 2,400 + 280) = \$199,000 per instance! 
To put this figure in perspective, we consider two alternatives---the estimate without adopting the technology and the estimate assuming reconfiguration addresses the false positives. 

Without this technology, if the lone attack per required 10m for detection and containment, then our cost estimate is simply \$480 $+  f(10)$ / month / instance = \$12,905 / month / instance. 
Applying the GL $1/e$ rule of thumb, we see optimal security costs should be below \$12,905/$e$ = \$4,747 per month per instance or \$56,970 per year per instance.

In the more interesting reconfiguration scenario, we note that  the false positives found in testing were occurring from legitimate network scanning appliances tripping the GraphPrints detector. 
Common practice for handling such false positives involves continually tuning tools \cite{DBLP:journals/corr/abs-1812-02867, sundaramurthy2016turning}. 
As we included the labor costs for monthly reconfiguration, it is reasonable to assume that each such false positive would occur one time, then reconfiguration would prevent the same alert. 
In this case, the there would be only the lone, first false positives in the testing window (Figure \ref{fig:graphprints-network-results}), so our false positive rate drops to 0.56\%. 
With this false positive rate, we incur 86,400 events $\times$ 0.56\% = 14 false positives per month per instance, for a cost of 14 $\times $ \$80 = \$1,120. 
Now the cost for adopting the GraphPrints technology is the initial \$9,185 server cost plus \$ (1,120 + 2,400 + 280) = \$3,800/month/instance. We note that this is indeed below the GL upper bound. 
Neglecting initial costs, the technology promises a yearly savings to customers of 12 months $\times$ \$(12,905 - \$3,800)/month = \$109,260. 
With the initial costs for hardware and configuration included, we see operations will save about \$100,000 / year.

From the point of a researcher, such an analysis is enlightening, as it allows quantitative reasoning about the impact of the true/false positive analysis and resource requirements. 
Further, it gives a single metric to optimize when, for example, deciding a threshold for detection. 
From a SOC's perspective, provided the numbers above are reasonable estimates, the conclusion is clear\textemdash if false positives can be mostly eliminated with one-off reconfiguration tuning, then this is a good investment; if not, then this is a terrible investment. 
We recommend a testing period to give a much more informed decision on both the figures estimated above and the final decision. 
Finally, from the perspective of the vendor, such an estimate can help dial in their yearly subscription fees. 
Yearly cost to use this technology are \$9,185 (server cost) $+ 12\times 3,800 $ = \$54,785 plus subscription fees.  
If a subscription is required per instance (scales with $n$), then annual subscription cost can be bounded above by the GL bound minus the operational costs, that is, they should be less than \$56,970 - 54,785 =  \$2,185 to keep total costs under the GL rule of thumb.

%% file: 90-conclusion.tex
\section{Conclusion}


Useful security metrics are important for estimating the efficacy of new products or new technologies, important for evaluating red team or competition events, and important for organizations which must weigh the cost verses benefit of security practices.
As we have described, each of those three areas have developed their own generally accepted metrics within their topic areas, but these have been focused too narrowly, and cannot easily be applied from one area to another. 
For example, it is currently difficult to take the statistical metrics from researcher testing of an IDS and estimate the impact on a specific organization. 
In this paper, we have proposed a unified approach, which generalizes and combines the traditional metrics in these areas in a flexible framework by comprehensively modeling the various costs involved. 
This provides a configurable cost model that balances accuracy, timeliness, and resource use. 
Moreover, it is easy to interpret and analyze. 
To illustrate the efficacy of the new model, we tune it to be used as the scoring procedure for an upcoming IARPA IDS competition, and use simulated attack/defense scenarios to test the efficacy of the cost framework. 
Our results support that a balance of accuracy, response time, and resource use are promoted by the model. 
Finally, we exhibit the use of this new model to evaluate a new security tool from multiple points of view,
specifically the researcher, the SOC (client), and the vendor. 
Our results show the model can provide clear and actionable insights from each. 

%% file: 99-acks.tex
\section*{Acknowledgements}
The authors would like to thank the many reviewers who have helped polish this document. 
Special thanks to Kerry Long for his insights and guidance during our authorship of this paper, to Miki Verma, Dave Richardson, Brian Jewell, and Jason Laska for helpful discussions, and to the many SOC operators who provided consultation in the preparation of this manuscript.

The research is based upon work supported by the Office of the Director of National Intelligence (ODNI), Intelligence Advanced Research Projects Activity (IARPA), via the Department of Energy (DOE) under contract  D2017-170222007. 
The views and conclusions contained herein are those of the authors and should not be interpreted as necessarily representing the official policies or endorsements, either expressed or implied, of the ODNI, IARPA, or the U.S. Government. 
The U.S. Government is authorized to reproduce and distribute reprints for Governmental purposes notwithstanding any copyright annotation thereon. 